\documentclass{aa}
\usepackage[utf8]{inputenc}
\usepackage{natbib}
\usepackage{graphicx}
\usepackage{amsmath}
\usepackage{gensymb}
\usepackage{subcaption}
\usepackage{dblfloatfix}
\usepackage{makecell}
\usepackage{ragged2e}
\usepackage[colorlinks=true,allcolors=blue]{hyperref}
\usepackage[scientific-notation=true]{siunitx}
\usepackage{xcolor}
\usepackage{mhchem}
\usepackage{longtable}
\usepackage{lscape}

\title{Intermediate mass T~Tauri disk masses and a comparison to their Herbig disk descendants}

\author{L. M. Stapper\inst{\ref{inst1}, \ref{inst1.2}} \and M. R. Hogerheijde\inst{\ref{inst1}, \ref{inst2}} \and E. F. van Dishoeck\inst{\ref{inst1},\ref{inst3}} \and M. Vioque \inst{\ref{inst4}} \and J. P. Williams\inst{\ref{inst5}} \and C. Ginski\inst{\ref{inst6}}}

\institute{Leiden Observatory, Leiden University, PO Box 9513, 2300 RA Leiden, The Netherlands \label{inst1} \and Max-Planck-Institut für Astronomie, Königstuhl 17, 69117 Heidelberg, Germany \\e-mail: \texttt{lustapper@mpia.de} \label{inst1.2} \and Anton Pannekoek Institute for Astronomy, University of Amsterdam, PO Box 94249, 1090 GE, Amsterdam, The Netherlands \label{inst2} \and Max-Planck-Institut für Extraterrestrische Physik, Giessenbachstrasse 1, 85748 Garching, Germany \label{inst3} \and European Southern Observatory, Karl-Schwarzschild-Strasse 2, 85748 Garching, Germany \label{inst4} \and Institute for Astronomy, University of Hawaii, Honolulu, Hawaii 96822 \label{inst5} \and School of Natural Sciences, Center for Astronomy, University of Galway, Galway, H91 CF50, Ireland \label{inst6}}

\date{\today}

\abstract
{ 
Herbig disks are the prime sites for the formation of massive exoplanets, so looking into the precursors of these disks can give us clues in planet formation timescales. The precursors of Herbig stars are called Intermediate Mass T~Tauri (IMTT) stars, which have spectral types later than F, but stellar masses between 1.5 and 5~$M_\odot$, and will eventually become Herbig stars with spectral types of A and B.
}
{ 
The aim of this work is to obtain the dust and gas masses and radii of all IMTT disks with ALMA archival data. The obtained disk masses are then compared to Herbig disks and T~Tauri disks and the obtained disks sizes to those of Herbig disks.
}
{ 
ALMA Band 6 and 7 archival data are obtained for 34 IMTT disks with continuum observations, 32 of which have at least \ce{^12CO}, \ce{^13CO}, or \ce{C^18O} observations although most of them at quite shallow integrations. The disk integrated flux together with a stellar luminosity scaled disk temperature are used to obtain a total disk dust mass by assuming optically thin emission. Using thermochemical Dust And LInes (DALI) models from previous work, we additionally obtain gas masses of 10/35 of the IMTT disks based on the CO isotopologues. From the disk masses and sizes cumulative distributions are obtained.
}
{ 
The IMTT disks in this study have the same dust mass and radius distributions as Herbig disks. The dust mass of the IMTT disks is higher compared to that of the T~Tauri disks, as is also found for the Herbig disks. No differences in dust mass are found for group~I versus group~II disks, in contrast to Herbig disks. The disks for which a gas mass could be determined show similar high mass disks as for the Herbig disks. Comparing the disk dust and gas mass distributions to the mass distribution of exoplanets shows that there also is not enough dust mass in disks around intermediate mass stars to form the massive exoplanets. On the other hand there is more than enough gas to form the atmospheres of exoplanets.
}
{ 
We conclude that the sampled IMTT disk population is almost indistinguishable compared to Herbig disks, as their disk masses are the same, even though these are younger objects. Based on this, we conclude that planet formation is already well on its way in these objects, and thus planet formation should start early on in the lifetime of Herbig disks. Combined with our findings on the group~I and group~II disks, we conclude that most disks around intermediate mass pre-main sequence stars converge quickly to small disks unless prevented by a massive exoplanet.
}


\begin{document}

\maketitle

\defcitealias{Stapper2022}{S22}
\defcitealias{Stapper2024}{S24}

\begin{table*}
\caption{Source parameters used to calculate the dust masses, used for Keplerian masking, and the obtained dust and gas radii.}
\tiny\centering
\resizebox{\textwidth}{!}{\begin{tabular}{l|rrccccccccccl}
\hline\hline
\makecell{Name \\ \hspace{1mm}} & \makecell{RA \\ (h:m:s)} & \makecell{Dec \\ (deg:m:s)} & \makecell{Dist. \\ (pc)} & \makecell{$M_\star$ \\ ($M_\odot$)} & \makecell{$L_\star$ \\ ($L_\odot$)}  & \makecell{Age \\ (Myr)} & \makecell{Sp.Tp. \\ \hspace{1mm}} & \makecell{Group \\ \hspace{1mm}} & \makecell{V$_\text{sys}$ \\ (km s$^{-1}$)} & \makecell{V$_\text{int}$ \\ (km s$^{-1}$)} & \makecell{inc \\ ($\degree$)} & \makecell{PA \\ ($\degree$)} & \makecell{Ref. \\ \hspace{1mm}} \\ \hline
AK Sco & 16:54:44.8 & -36:53:19.0 & 140 & 1.52 & 6.94 & 8.57 & F5 & I & 5.4 & 2.5 & 109 & 128 & 1, a \\
Ass ChaT2-21 & 11:06:15.3 & -77:21:56.7 & 165 & 2.09 & 10.37 & 3.23 & G5 & debris & -- & -- & -- & -- & 2 \\
Ass ChaT2-54 & 11:12:42.6 & -77:22:22.9 & 203 & 1.92 & 5.03 & 3.06 & G8 & debris & -- & -- & -- & -- & 3 \\
BE Ori & 05:37:00.1 & -06:33:27.3 & 392 & 1.86 & 7.95 & 4.55 & G3 & II & -- & -- & -- & -- & 2 \\
Brun 656 & 05:35:21.3 & -05:12:12.7 & 467 & 2.72 & 27.4 & 1.53 & G2 & debris & -- & -- & -- & -- & 2 \\
CO Ori & 05:27:38.2 & +11:25:39.1 & 400 & 2.97 & 43.82 & 1.48 & F7 & II & -- & -- & -- & -- & 4 \\
CR Cha & 10:59:06.9 & -77:01:40.3 & 186 & 1.62 & 3.72 & 1.51 & K4 & II & 5.2 & 2.5 & 31 & -36 & 5, b \\
CV Cha & 11:12:27.6 & -76:44:22.3 & 192 & 1.85 & 4.63 & 3.44 & K0 & II & -- & -- & -- & -- & 2 \\
DI Cha & 11:07:20.6 & -77:38:07.3 & 190 & 1.95 & 9.82 & 4.08 & G2 & II & -- & -- & -- & -- & 2 \\
GW Ori & 05:29:08.4 & +11:52:12.7 & 398 & 3.02 & 35.47 & 1.00 & G5 & I & 13.8 & 2.5 & 37 & -176 & 6, c \\
Haro 1-6 & 16:26:03.0 & -24:23:36.6 & 134 & 2.04 & 12.78 & 3.50 & G1 & I & -- & -- & -- & -- & 2 \\
HBC 442 & 05:34:14.2 & -05:36:54.2 & 382 & 1.9 & 13.2 & 4.75 & F8 & II & -- & -- & -- & -- & 7 \\
HBC 502 & 05:46:07.9 & -00:11:56.9 & 411 & 2.01 & 10.71 & 0.53 & K3 & I & -- & -- & -- & -- & 2 \\
HD 34700 & 05:19:41.4 & +05:38:42.8 & 353 & 2.46 & 24.34 & 2.41 & G0 & I & 5.1 & 4.5 & 0 & 0 & 2 \\
HD 35929 & 05:27:42.8 & -08:19:38.5 & 384 & 3.25 & 80.78 & 1.14 & F2 & II & -- & -- & -- & -- & 8 \\
HD 135344B & 15:15:48.4 & -37:09:16.4 & 135 & 1.5 & 8.09 & 9.07 & F8 & I & 7.1 & 1.5 & 10 & -62 & 2, d \\
HD 142527 & 15:56:41.9 & -42:19:23.7 & 157 & 2.06 & 19.31 & 3.80 & F6 & I & 3.6 & 1.5 & 27 & 26 & 9, e \\
HD 142666 & 15:56:40.0 & -22:01:40.4 & 148 & 1.62 & 11.81 & 8.37 & A8 & II & 4.1 & 1.5 & 62 & 18 & 2, f \\
HD 144432 & 16:06:57.9 & -27:43:10.1 & 155 & 1.64 & 12.5 & 8.14 & A9 & II & -- & -- & -- & -- & 10 \\
HD 288313 & 05:54:03.0 & +01:40:22.1 & 418 & 3.47 & 38.47 & 0.52 & K2 & -- & -- & -- & -- & -- & 2 \\
HD 294260 & 05:36:51.3 & -04:25:40.0 & 407 & 1.68 & 8.74 & 6.72 & G1 & II & 12.0 & 2.5 & 0 & 0 & 11 \\
HQ Tau & 04:35:47.3 & +22:50:21.4 & 158 & 1.85 & 4.63 & 3.44 & K0 & II & -- & -- & -- & -- & 2 \\
HT Lup & 15:45:12.8 & -34:17:31.0 & 154 & 1.8 & 5.95 & 0.95 & K3 & II & 5.4 & 2.5 & 48 & 14 & 2, g \\
LkHA 310 & 05:47:11.0 & +00:19:14.8 & 422 & 1.87 & 6.8 & 4.25 & G6 & I & -- & -- & -- & -- & 2 \\
LkHA 330 & 03:45:48.3 & +32:24:11.8 & 308 & 1.93 & 14.39 & 4.66 & F7 & I & 9.0 & 0.8 & 28 & -49 & 2, h \\
PDS 156 & 18:27:26.1 & -04:34:47.6 & 398 & 2.61 & 21.12 & 1.80 & G5 & II & 13.0 & 2.5 & 0 & 0 & 2 \\
PDS 277 & 08:23:11.8 & -39:07:01.5 & 343 & 1.59 & 9.68 & 8.66 & F3 & I & 1.6 & 2.5 & 0 & 0 & 2 \\
PR Ori & 05:36:24.8 & -06:17:30.6 & 408 & 2.5 & 10.83 & 1.05 & K1 & I/II & -- & -- & -- & -- & 2 \\
RY Ori & 05:32:09.9 & -02:49:46.8 & 365 & 1.69 & 9.01 & 6.65 & F7 & II & -- & -- & -- & -- & 7 \\
RY Tau & 04:21:57.4 & +28:26:35.1 & 134 & 1.69 & 11.97 & 4.29 & K1 & II & 6.6 & 2.5 & 65 & -23 & 6, i \\
EM* SR 21 & 16:27:10.3 & -24:19:13.1 & 138 & 1.64 & 7.03 & 6.71 & G1 & I & 2.75 & 2.5 & 15 & -16 & 12, j \\
SU Aur & 04:56:00.4 & +30:34:01.1 & 158 & 2.22 & 12.75 & 2.92 & G4 & I & 5.8 & 1.5 & 53 & 123 & 2, k \\
SW Ori & 05:34:15.8 & -06:36:04.7 & 376 & 1.51 & 3.28 & 7.10 & G8 & I & -- & -- & -- & -- & 2 \\
T Tau & 04:21:59.4 & +19:32:06.2 & 144 & 1.94 & 8.88 & 4.01 & K0 & I & -- & -- & -- & -- & 6 \\
UX Tau & 04:30:04.0 & +18:13:49.2 & 139 & 2.34 & 8.91 & 1.26 & G8 & I & 5.4 & 2.5 & 40 & -167 & 2, l \\
\hline
\end{tabular}}
\label{tab:disk_parameters}
\tablefoot{Object coordinates, distances, stellar masses, luminosities, ages, and groups are taken from \citet{Valegard2021} and references therein. See \citet{Valegard2021} for the uncertainties on these values. The spectral types are taken from: (1) \citet{Houk1982}, (2) \citet{Valegard2021}, (3) \citet{Daemgen2013}, (4) \citet{Mora2001}, (5) \citet{Torres2006}, (6) \citet{Herbig1977}, (7) \citet{Manoj2006}, (8) \citet{Houk1999}, (9) \citet{Houk1978}, (10) \citet{Houk1982}, (11) \citet{Rydgren1984}, (12) \citet{Suarez2006}. The inclinations and position angles are taken from: (a) \citet{Czekala2015}, (b) \citet{Kim2020}, (c) \citet{Bi2020}, (d) \citet{Cazzoletti2019}, (e) \citet{Kataoka2016}, (f) \citet{Huang2018}, (g) \citet{Kurtovic2018}, (h) \citet{Pinilla2022}, (i) \citet{Long2018}, (j) \citet{Perez2014}, (k) \citet{Ginski2021}, (l) \citet{Francis2020}. For the disks with no detection no parameters are given, as indicated by the dashed lines. For the disks with a zero degree inclination and position angle no values were found in the literature, but a mask was applied. The velocities of the Keplerian masks were determined by fitting by eye, such that all emission was enclosed.}
\end{table*}

\begin{sidewaystable*}
\caption{Continuum and CO line ($J=2-1$ or $J=3-2$) fluxes as well as dust and gas radii.}
\tiny\centering
\resizebox{\textwidth}{!}{\begin{tabular}{l|ccc|ccc|ccc|ccc|cc}
\hline\hline
 & \multicolumn{3}{c|}{Continuum} & \multicolumn{3}{c|}{\ce{^12CO}} & \multicolumn{3}{c|}{\ce{^13CO}} & \multicolumn{3}{c|}{\ce{C^18O}} & & \\ \hline
\makecell{Name \\ \hspace{1mm}} & \makecell{Flux \\ (mJy)} & \makecell{R$_\text{dust, 68\%}$ \\ (au)} & \makecell{R$_\text{dust, 90\%}$ \\ (au)} & \makecell{Flux \\ (Jy~km~s$^{{-1}}$)} & \makecell{R$_\text{gas, 68\%}$ \\ (au)} & \makecell{R$_\text{gas, 90\%}$ \\ (au)} & \makecell{Flux \\ (Jy~km~s$^{{-1}}$)} & \makecell{R$_\text{gas, 68\%}$ \\ (au)} & \makecell{R$_\text{gas, 90\%}$ \\ (au)} & \makecell{Flux \\ (Jy~km~s$^{{-1}}$)} & \makecell{R$_\text{gas, 68\%}$ \\ (au)} & \makecell{R$_\text{gas, 90\%}$ \\ (au)} & \makecell{$M_\text{dust}$ \\ ($M_\oplus$)} & \makecell{Log$_{10}$($M_\text{gas}$) \\ ($M_\odot$)} \\ \hline
AK Sco & 26.14 & 42±2 & 53±2 & 2.18±0.05 & 78±4 & 114±5 & 0.68±0.04 & 66±6 & 97±18 & 0.27±0.02 & 52±5 & 69±11 & 6.0±0.6 & -2.6±1.2 \\
Ass ChaT2-21 & <1.21$^{*}$\hspace{-4.4pt}	 & -- & -- & <0.29$^{*}$\hspace{-4.4pt}	 & -- & -- & -- & -- & -- & -- & -- & -- & <0.1 & --  \\
Ass ChaT2-54 & <3.16$^{*}$\hspace{-4.4pt}	 & -- & -- & <0.89$^{*}$\hspace{-4.4pt}	 & -- & -- & <5.06$^{*}$\hspace{-4.4pt}	 & -- & -- & <5.20$^{*}$\hspace{-4.4pt}	 & -- & -- & <0.5 & --  \\
BE Ori & 0.94 & <286 &<373 & <0.33	 & -- & -- & <0.39	 & -- & -- & <0.31	 & -- & -- & 1.7±0.2 & --  \\
Brun 656 & <1.68	 & -- & -- & -- & -- & -- & -- & -- & -- & -- & -- & -- & <3.0 & --  \\
CO Ori & <7.78	 & -- & -- & <2.09	 & -- & -- & <2.51	 & -- & -- & <1.85	 & -- & -- & <10.7 & --  \\
CR Cha & 139.47 & 63±3 & 88±3 & 2.22±0.04 & 158±27 & 249±27 & 0.74±0.05 & 150±28 & 233±111 & 0.29±0.02 & 139±28 & 201±28 & 69.7±7.0 & -1.7±0.8 \\
CV Cha & 56.70$^{*}$\hspace{-4.4pt} & <76 &<119 & -- & -- & -- & <4.99	$^{*}$\hspace{-4.4pt} & -- & -- & <4.51$^{*}$\hspace{-4.4pt}	 & -- & -- & 10.4±1.0 & --  \\
DI Cha & 23.76$^{*}$\hspace{-4.4pt} & <73 &<100 & -- & -- & -- & <4.48$^{*}$\hspace{-4.4pt}	 & -- & -- & <1.80$^{*}$\hspace{-4.4pt}	 & -- & -- & 3.4±0.3 & --  \\
GW Ori & 200.37 & 286±12 & 381±12 & 51.09±0.06 & 778±16 & 1032±16 & 6.28±0.07 & 576±16 & 805±20 & 1.30±0.03 & 418±16 & 559±20 & 241.9±27.3 & -1.1±0.4 \\
Haro 1-6 & 3.22 & <108 &<155 & <0.88	 & -- & -- & <0.99	 & -- & -- & <0.44	 & -- & -- & 0.6±0.1 & --  \\
HBC 442 & <9.29	 & -- & -- & <2.41	 & -- & -- & <2.58	 & -- & -- & <2.42	 & -- & -- & <16.2 & --  \\
HBC 502 & <4.57	 & -- & -- & -- & -- & -- & -- & -- & -- & -- & -- & -- & <7.9 & --  \\
HD 34700 & 10.51 & <189 &<244 & 6.65±0.06 & 301±38 & 421±38 & <1.20	 & -- & -- & <0.91	 & -- & -- & 10.7±1.1 & --  \\
HD 35929 & 0.22 & <12 &<17 & <0.04	 & -- & -- & <0.04	 & -- & -- & <0.03	 & -- & -- & 0.23±0.03 & --  \\
HD 135344B & 539.69$^{*}$\hspace{-4.4pt} & 81±10 & 109±10 & 20.07±0.18$^{*}$\hspace{-4.4pt} & 158±10 & 219±10 & 8.86±0.08$^{*}$\hspace{-4.4pt} & 119±9 & 171±9 & 3.55±0.08$^{*}$\hspace{-4.4pt} & 86±9 & 125±9 & 35.8±3.7 & -1.7±0.4 \\
HD 142527 & 1048.15 & 208±9 & 236±9 & 27.14±0.10 & 542±30 & 793±30 & 11.88±0.07 & 334±31 & 491±31 & 3.91±0.04 & 271±31 & 368±31 & 231.2±23.4 & -0.8±0.2 \\
HD 142666 & 118.55 & 40±1 & 52±1 & 4.05±0.03 & 136±6 & 186±6 & 1.51±0.06 & <182 &<260 & 0.64±0.04 & <164 &<236 & 26.0±2.6 & -1.5±0.7 \\
HD 144432 & 44.92 & <587 &<856 & <2.25	 & -- & -- & <1.59	 & -- & -- & <1.21	 & -- & -- & 13.1±1.3 & --  \\
HD 288313 & 1.03 & <99 &<166 & <0.15	 & -- & -- & <0.18	 & -- & -- & <0.16	 & -- & -- & 1.6±0.2 & --  \\
HD 294260 & 41.53 & <123 &<181 & 1.72±0.03 & <233 &<348 & -- & -- & -- & -- & -- & -- & 74.4±7.9 & --  \\
HQ Tau & 3.94 & <22 &<30 & -- & -- & -- & <0.18	 & -- & -- & <0.13	 & -- & -- & 1.6±0.2 & --  \\
HT Lup & 62.54 & 19±1 & 25±1 & 4.78±0.04 & 111±2 & 141±2 & <0.24	 & -- & -- & <0.18	 & -- & -- & 18.0±1.8 & --  \\
LkHA 310 & 14.55 & <109 &<157 & <0.18	 & -- & -- & <0.16	 & -- & -- & <0.15	 & -- & -- & 36.7±4.2 & --  \\
LkHA 330 & 52.33 & 123±5 & 141±5 & 2.00±0.12 & 219±23 & 337±54 & 1.48±0.08 & 194±14 & 261±29 & 0.63±0.04 & 117±7 & 143±12 & 47.2±5.3 & -1.2±0.5 \\
PDS 156 & 9.94 & <90 &<129 & 0.35±0.03 & <164 &<273 & <0.06	 & -- & -- & <0.05	 & -- & -- & 16.2±1.7 & --  \\
PDS 277 & 39.48 & <114 &<168 & 0.62±0.04 & <217 &<322 & <0.72	 & -- & -- & <0.67	 & -- & -- & 48.8±5.0 & --  \\
PR Ori & 21.82 & <344 &<510 & -- & -- & -- & -- & -- & -- & -- & -- & -- & 41.7±4.3 & --  \\
RY Ori & <22.49	 & -- & -- & <2.58	 & -- & -- & <2.50	 & -- & -- & <2.16	 & -- & -- & <40.0 & --  \\
RY Tau & 207.91 & 51±1 & 65±1 & 7.89±0.20 & 171±7 & 237±10 & 1.85±0.09 & 114±7 & 156±14 & 0.46±0.05 & 86±10 & 112±21 & 37.1±23.9 & -3.3±0.4  \\
SR 21 & 88.20 & 61±3 & 69±3 & 0.53±0.02 & 28±4 & 35±4 & 0.50±0.02 & 31±4 & 42±4 & 0.38±0.02 & 34±4 & 47±4 & 23.8±2.4 & -1.2±0.6 \\
SU Aur & 16.73 & <58 & <100 & 13.52±0.13 & 255±11 & 346±11 & 0.91±0.13 & 211±65 & 320±141 & -- & -- & -- & 4.9±0.5 & -- \\
SW Ori & 7.46 & <285 &<396 & <0.38	 & -- & -- & <0.44 & -- & -- & <0.34	 & -- & -- & 15.7±1.6 & --  \\
T Tau & 171.62 & 16±1 & 22±1 & -- & -- & -- & -- & -- & -- & -- & -- & -- & 47.3±4.8 & --  \\
UX Tau & 64.54 & 49±7 & 64±7 & 8.46±0.12 & 216±9 & 453±24 & 2.10±0.06 & 94±9 & 152±13 & 0.96±0.04 & 82±9 & 132±20 & 13.7±1.4 & -1.4±0.6 \\
\hline
\end{tabular}}
\label{tab:disk_fluxes_radii}
\tablefoot{For AK~Sco, HD~135344B, HD~142526, and HD~142666 the fluxes, radii, and gas masses  were taken from \citet{Stapper2022} and \citet{Stapper2024}. The dust masses were recalculated using the stellar luminosities of \citet{Valegard2021}. All fluxes with an asterisk are for the $J=3-2$ transition, the other fluxes are for the $J=2-1$ transition. A non-detection is indicated by a dash. On all fluxes there is an additional 10\% calibration error.}
\end{sidewaystable*}

\section{Introduction}
\label{sec:introduction}
Herbig disks are planet-forming disks around pre main-sequence stars called Herbig stars characterized by spectral types of B, A, and F, and stellar masses of 1.5~$M_\odot$ to 10~$M_\odot$ \citep[e.g.,][]{Herbig1960, Brittain2023}. These disks are the precursors of famous directly imaged planetary systems such as HR~8799 \citep{Marois2008, Marois2010}, $\beta$~Pic \citep{Lagrange2010}, and 51~Eri \citep{Chauvin2017}, and host some of the first kinematically detected planets \citep{Pinte2018, Izquierdo2023}. How these planets have formed is not known, but we do know from exoplanet statistics that the prevalence of giant planets is highest around intermediate mass stars \citep[e.g.,][]{Johnson2007, Johnson2010, Nielsen2019, Fulton2021}. Comparing dust structures and planet occurrence rates indeed shows a positive correlation between the two, and an increase in the prevalence of dust structures with stellar mass \citep{vanderMarel2021}. Hence, Herbig disks are an important piece of the planet formation puzzle.

\begin{figure}
    \centering
    \includegraphics[width=0.5\textwidth]{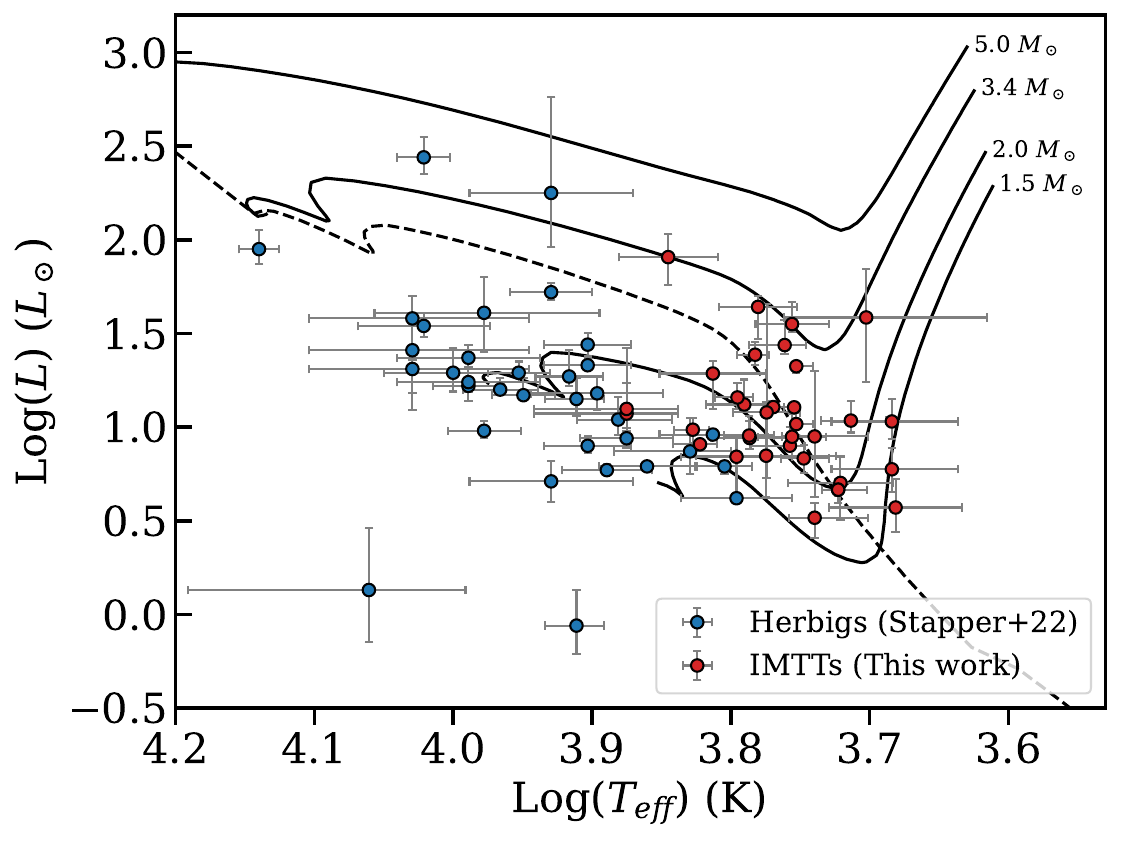}
    \caption{An HR-diagram of the Herbig star sample of \citetalias{Stapper2022}, and the compiled IMTT sample in this work. The effective temperatures and luminosities are taken from \citet{Vioque2018} and \citet{Valegard2021} for the Herbig stars and IMTT stars respectively. Pre-main sequence tracks of different intermediate mass stars, and an isochrone of 2.5~Myr as the dashed line, are shown from \citet{Marigo2017}.}
    \label{fig:hr_diagram}
\end{figure}

\begin{figure*}[t]
    \centering
    \includegraphics[width=0.8\textwidth]{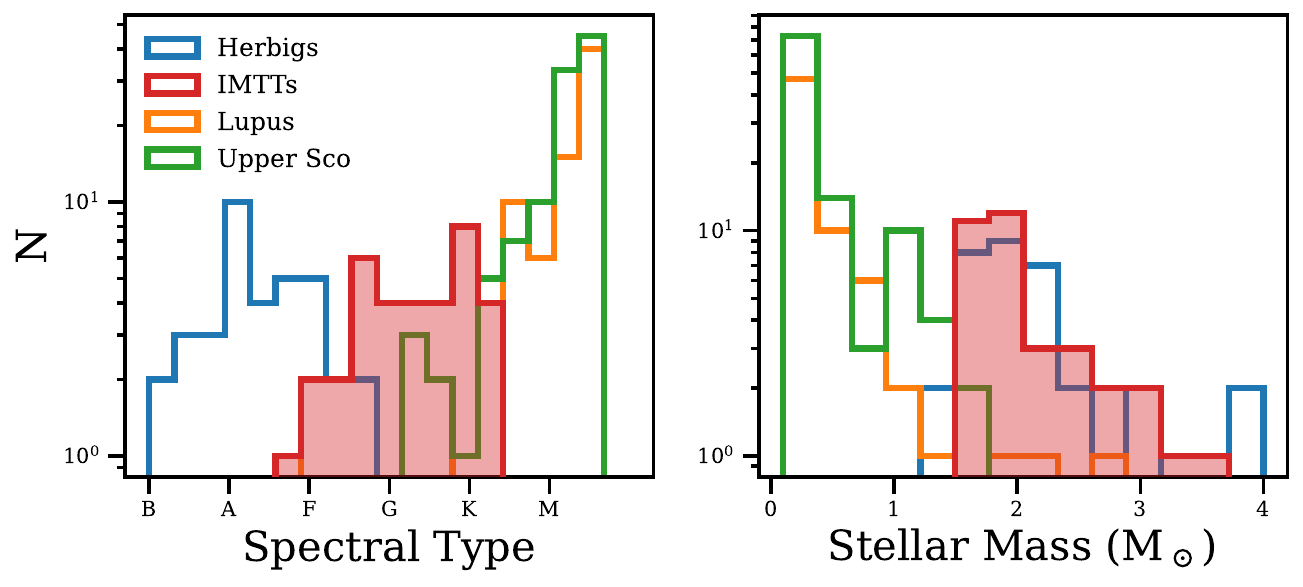}
    \caption{Histogram of the spectral types (left panel) and stellar masses (right panel) of the IMTTs in this work \citep[from][]{Valegard2021} compared to the Herbig star sample of \citetalias{Stapper2022} and references therein, and the surveys of Lupus \citep{Ansdell2016} and Upper~Sco \citep{Barenfeld2016}.}
    \label{fig:sptp_hist}
\end{figure*}

However, Herbig disks are generally quite old (a median age of 6~Myr for Herbig Ae stars, \citealt{Vioque2018}), many showing structures indicative of planet formation making these some of the most famous and best investigated disks (e.g., HD~100546, HD~163296, MWC~480; \citealt{Fedele2017}; \citealt{Teague2018}; \citealt{Oberg2021}; \citealt{Booth2023}). Planets may already form early on in the lifetime of the disk, as there is an apparent lack of solids in class~II disks to form the observed family of giant exoplanets \citep{Manara2018, Tychoniec2020}, and structures are already visible in earlier stages \citep{ALMApartnership2015, SeguraCox2020}. While the recent ALMA large program eDisk \citep{Ohashi2023} does not show as many substructures in class~0 and class~I objects as expected, possibly because these structures have not formed yet or have high continuum optical depth, planetesimals, i.e., planets' building blocks, are likely to already form in these younger objects \citep[e.g.,][]{Drazkowska2018}. Hence, these younger objects can still give us insights into planet formation timescales.

In total around 380 Herbig stars are known \citep{Vioque2018, Wichittanakom2020, GuzmanDiaz2021, Vioque2022}. In addition, around 1500 more Herbig candidates have been identified by \citet{Vioque2020}, so the total population size is expected to be much larger. A recent compilation of all available ALMA archival data out to Orion ($\sim450$~pc) has shown that these Herbig disks are more massive in dust mass than the disks around their lower stellar mass counterparts (with spectral type F and later, and a stellar mass of $\lesssim1.5$~$M_\odot$) called T~Tauri stars \citep{Stapper2022}. This could naturally result in the higher prevalence of giant exoplanets around higher mass stars \citep[e.g.,][]{Johnson2010}. Moreover, work by \citet{Stapper2024} indicates that the Herbig disks are much warmer compared to T~Tauri disks, causing less CO freeze-out and reprocessing in Herbig disks, consistent with thermo-chemical models \citep{Bosman2018}. This makes CO a viable mass tracer in Herbig disks. 

However, due to the relatively horizontal pre main-sequence evolutionary tracks in the Hertzsprung-Russell diagram, there are intermediate mass objects which were not part of the analysis of \citet{Stapper2022, Stapper2024} which have the same stellar mass as Herbig stars but a spectral type later than F. These are called Intermediate mass T~Tauri (IMTT) stars, the precursors of Herbig stars, with spectral types of F to K3 and stellar masses between 1.5 and 5~$M_\odot$, which look like T~Tauri stars but have an intermediate mass star \citep{Calvet2004, Valegard2021}. This is shown in the Hertzsprung-Russel diagram in Fig~\ref{fig:hr_diagram} in which the difference in effective temperatures can be seen between the Herbig stars and IMTT stars. As a reference the 2.5~Myr isochrone is shown, which clearly indicates that the IMTT stars are younger than their Herbig star descendants, though the ages can be relatively uncertain. \citet{Valegard2021} has compiled a sample of 49 IMTTs based on optical photometry. All stars within 500~pc with spectral types ranging from F0 to K3, and with a stellar luminosity of at least 2.1~$L_\odot$ (i.e., $M_\star\geq1.5$~$M_\odot$) were selected. The resulting sample, while not as young as class~I objects, has a median age of 4~Myr, ranging from 0.3~Myr to 9~Myr, based on isochrones. We note that, like the Herbig disks, the IMTT disks are likely biased towards the high accretors and brightest disks (see, e.g., Figure~1 of \citealt{GrantStapper2023}). A millimeter study into this sample will give insight into the evolution of the dust and gas around these intermediate mass objects.

A significant fraction of the IMTTs compiled by \citet{Valegard2021} has ALMA millimeter observations. In this work we compile the available ALMA data and compare the obtained dust masses to previous works of Herbig disks \citep[][hereafter \citetalias{Stapper2022}]{Stapper2022}, and gas masses to Herbig disks \citep[][hereafter \citetalias{Stapper2024}]{Stapper2024}, and to T~Tauri disks \citep[e.g.,][]{Ansdell2016, Ansdell2017, Barenfeld2016, vanTerwisga2020, vanTerwisga2022, Manara2023}.

In Section~\ref{sec:data_selection_reduction} the data selection and reduction procedures are set out. In Section~\ref{sec:model_setup} we explore if new models are needed compared to the DALI \citep{Bruderer2012, Bruderer2013} thermochemical models run for Herbig disks in \citetalias{Stapper2024}, which can then be used to determine the gas masses from the CO observations. In Section~\ref{sec:results} we show the resulting continuum and gas images in \S\ref{subsec:continuum_images} and \S\ref{subsec:gas_images} respectively, obtain a dust mass and dust radius distribution and compare these to previous works in \S\ref{subsec:dust_masses}, and obtain the gas masses and radii in \S\ref{subsec:gas_masses} and compare these to previous works. We discuss these results in Section~\ref{sec:discussion}, in the context of the \citet{Meeus2001} groups and the evolution of disks around intermediate mass stars in \S\ref{subsec:evolution}, and discuss the implications on planet formation around intermediate mass stars in \S\ref{subsec:planet_formation}. Lastly, Section~\ref{sec:conclusion} summarizes our conclusions.

\section{Data selection and reduction}
\label{sec:data_selection_reduction}
Using the IMTTs from \citet{Valegard2021}, we obtained all publicly available Band~6 and Band~7 data on the ALMA archive\footnote{\url{https://almascience.eso.org/asax/}} containing continuum, \ce{^12CO}, \ce{^13CO}, and \ce{C^18O} observations. There is some overlap between the Herbig disks and IMTTs from \citet{Vioque2018} and \citet{Valegard2021}, the former have been used in \citetalias{Stapper2022} for Herbig dust masses. These are AK~Sco, HD~135344B, HD~142527, and HD~142666, see Table~\ref{tab:disk_parameters} for their spectral types. We use the data of these objects as presented in both \citetalias{Stapper2022} and \citetalias{Stapper2024}. Additionally we include recently made public ALMA data from the DESTINYS program (e.g., \citealt{Garufi2024}, \citealt{Ginski2024}, \citealt{Valegard2024}) with project code 2021.1.01705.S (P.I.: C. Ginski), and not yet published data with project code 2022.1.01155.S (P.I.: M. Vioque), and ACA data with project codes 2021.2.00005.S and 2022.1.01460.S (both P.I.: J. Williams). The resulting sample of IMTTs has spectral types ranging from A8 to K4. The histograms presented in Fig.~\ref{fig:sptp_hist} show that this range in spectral types overlaps with the late spectral type Herbig stars and the early spectral type T~Tauri stars, but that the range in stellar masses is the same as that of the Herbig stars. Also, see Figure~\ref{fig:hr_diagram} for a comparison on the HR-diagram between the Herbig stars and the IMTT stars of the sample in \citetalias{Stapper2022} and this work respectively.

Comparing the stellar parameters of the resulting sample of 35 IMTTs to the full sample of \citet{Valegard2021}, no significant differences regarding the stellar luminosity and mass are found. Using the Spitzer 30~$\mu$m fluxes from \citet{Valegard2021} as a measure of the presence of a disk, no significant differences are found either. See for a similar comparison for the Herbig disks between the sample of \citet{Stapper2022} and the full sample of \citet{Vioque2018} Appendix~D in \citet{Stapper2022}. No significant differences were found for the Herbig disks as well. Still, the samples themselves might be biased towards the highest accretors \citep{GrantStapper2023}, or biased towards the largest and brightest disks of the total population. But, as both the IMTTs and the Herbigs are biased in the same way, comparisons can still be made.


\begin{table*}
\caption{Observing details of the unpublished data.}
\tiny\centering
\resizebox{\textwidth}{!}{\begin{tabular}{ll|llclc}
\hline\hline
\makecell{Project Code \\ \hspace{1mm}} & \makecell{P.I. \\ \hspace{1mm}} & \makecell{Name \\ \hspace{1mm}} & \makecell{Date \\ \hspace{1mm}} & \makecell{Int. Time \\ (min.)} & \makecell{Calibrators \\ \hspace{1mm}} & \makecell{Array \\ \hspace{1mm}} \\ \hline
2021.1.01705.S & C. Ginski   & HD~294260       & 21/08/22, 02/09/22 & 2.6 & J0423-0120, J0532-0307 & 12m \\
               &             & HD~34700        & 21/08/22           & 1.4 & J0423-0120, J0527+0331 & 12m \\
               &             & PDS~277         & 31/08/22           & 2.4 & J1037-2934, J0828-3731 & 12m \\\hline
2022.1.01155.S & M. Vioque   & HD~288313~A     & 06/04/23, 15/06/23 & 7.3 & J0423-0120, J0541-0541 & 12m \\
               &             & LkH$\alpha$~310 & 06/04/23, 15/06/23 & 7.5 & J0423-0120, J0541-0541 & 12m \\
               &             & PDS~156         & 09/04/23           & 3.7 & J1924-2914, J1851+0035 & 12m \\\hline
2021.2.00005.S & J. Williams & CO~Ori          & 01/06/23           & 4.0 & J0423-0120, J0532+0732 & ACA \\
               &             & HD~34700        & 01/06/23           & 7.4 & J0423-0120, J0532+0732 & ACA \\
               &             & HBC~442         & 03/07/23           & 3.4 & J0423-0120, J0501-0159 & ACA \\
               &             & RY~Ori          & 03/07/23           & 3.4 & J0423-0120, J0501-0159 & ACA \\
               &             & PDS~277         & 28/08/22, 06/09/22 & 7.4 & J0538-4405, J0854+2006, J0501-0159 & ACA \\\hline
2022.1.01460.S & J. Williams & HD~144432       & 24/05/23           & 4.9 & J1924-2914, J1554-2704 & ACA \\
\hline
\end{tabular}}
\label{tab:unpublished_data}
\tablefoot{For the resulting spatial and velocity resolution, and the rms noise see Table~\ref{tab:project_codes}.}
\end{table*}

The observing details of the unpublished data can be found in Table~\ref{tab:unpublished_data}. In general the integration times are of the order of minutes, ranging from 1.4 minutes to 7.5 minutes. The data were taken in 2022 and 2023, almost all within a year of each other. As project 2021.1.01705.S only covers continuum and the \ce{^12CO} line, we supplement the data with other projects, where possible, to include the \ce{^13CO} and \ce{C^18O} lines. Specifically for HD~34700 and PDS~277 only the \ce{^13CO} and \ce{C^18O} data from 2021.2.00005.S are taken, and the \ce{^12CO} and continuum data are taken from the more sensitive observations of the 12m array.

The 12-m array data were phase self-calibrated based on the continuum data for multiple rounds up until the peak signal-to-noise did not improve compared to the previous round. This increased the peak signal-to-noise (S/N) of 12 of our targets (AK~Sco, CR~Cha, HD~135344~B, HD~142527, HD~294260, HT~Lup, PDS~156, PDS~277, RY~Tau, SR~21, SU~Aur, and UX~Tau). In most cases one to three rounds were done, beginning with a solution interval at ``inf'', then decreasing by factors of two, which increases the signal-to-noise ratio by a factor of 1.7 on average. After phase calibration, a single round of amplitude calibration was done as well, which was only applied to SU~Aur. The resulting calibration table was applied to the line spectral windows by using the \texttt{applycal} task. For the ACA data no self-calibration was done. In this work, for T~Tau, we only use the continuum data as the CO observations show very complex structures making it difficult to obtain a good measure of the disk mass and size. We refer for the T~Tau CO data to \citet{Rota2022}.

For the continuum data, the imaging was done using multifrequency synthesis. The spectral lines were imaged after subtracting the continuum using \texttt{uvcontsub}. Different velocity resolutions were used depending on the dataset. For the imaging we used \texttt{multiscale} using 0 (point source), 1, 2, 5, 10 and 15 times the size of the beam in pixels ($\sim5$ pixels) as the size of the scales. The last three scales were only used if the disk morphology allowed for it. Lastly, for the mosaic data of Brun~656 and HBC~502, we use the product data from the archive. See Table~\ref{tab:project_codes} for the resulting data parameters.

To obtain the disk integrated fluxes, we use the same method as described in \citetalias{Stapper2022} and \citetalias{Stapper2024} which uses an increasing aperture size to ascertain what the maximum amount of disk integrated flux is. This method also returns a size of the disk. The found fluxes and sizes are presented in Table~\ref{tab:disk_fluxes_radii}.

To obtain the dust masses, we follow previous population studies (see e.g., \citealt{Miotello2023} for an overview), using the relationship of \citet{Hildebrand1983} to directly relate the continuum emission to the dust mass, assuming optically thin emission,

\begin{equation}
    M_\text{dust} = \frac{F_\nu d^2}{\kappa_\nu B_\nu(T_\text{dust})}.
    \label{eq:mdust}
\end{equation}

Here, $F_\nu$ is the continuum flux as emitted by the dust in the disk at a distance $d$ to the object, and $\kappa_\nu$ is the dust opacity which is estimated as a power-law of the form $\kappa_\nu \propto \nu^\beta$, such that it equals 10~cm$^2$~g$^{-1}$ at a frequency of 1000~GHz \citep{Beckwith1990}. The power-law index $\beta$ is assumed to be equal to 1. $B_\nu$($T_\text{dust}$) is the value of the Planck function at a given dust temperature $T_\text{dust}$. The dust temperature is given by the relationship of \citet{Andrews2013}, which scales the dust temperature by the stellar luminosity in solar luminosities via

\begin{equation}
    T_\text{dust} = \text{25 K} \times \left( \frac{L_\star}{L_\odot} \right)^{1/4}.
\end{equation}

\begin{figure*}[t]
    \centering
    \includegraphics[width=\textwidth]{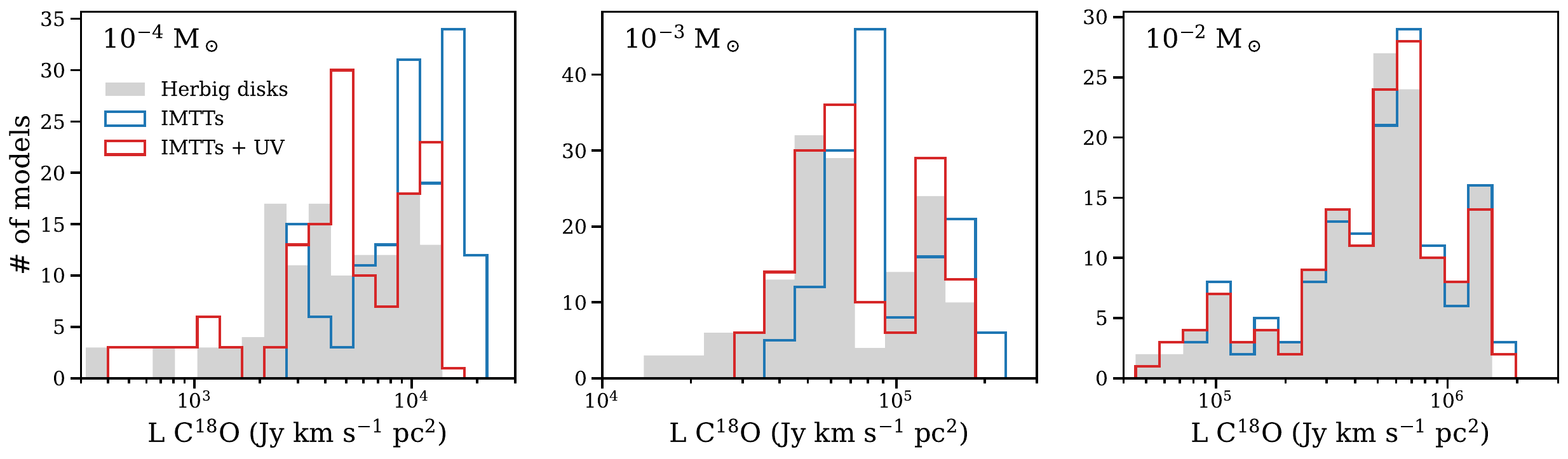}
    \caption{Histogram of the disk integrated \ce{C^18O} luminosities of models run by the thermochemical code DALI for three different disk masses, as indicated in the top left of each panel. A comparison is done between the models used by \citetalias{Stapper2024} (gray), and models for which the stellar effective temperature has been lowered to an IMTT appropriate value either with (red) or without (blue) accretion UV added. The lower effective temperature of the IMTT star increases the \ce{C^18O} luminosity for the lowest mass disks compared to the Herbig disk models, as there is less UV emission and therefore less photodissociation of CO. Adding the accretion UV increases the photodissociation of CO, moving the \ce{C^18O} luminosity back to Herbig disk levels.}
    \label{fig:IMTT_model_hists}
\end{figure*}

\begin{figure*}[t]
    \centering
    \includegraphics[width=0.9\textwidth]{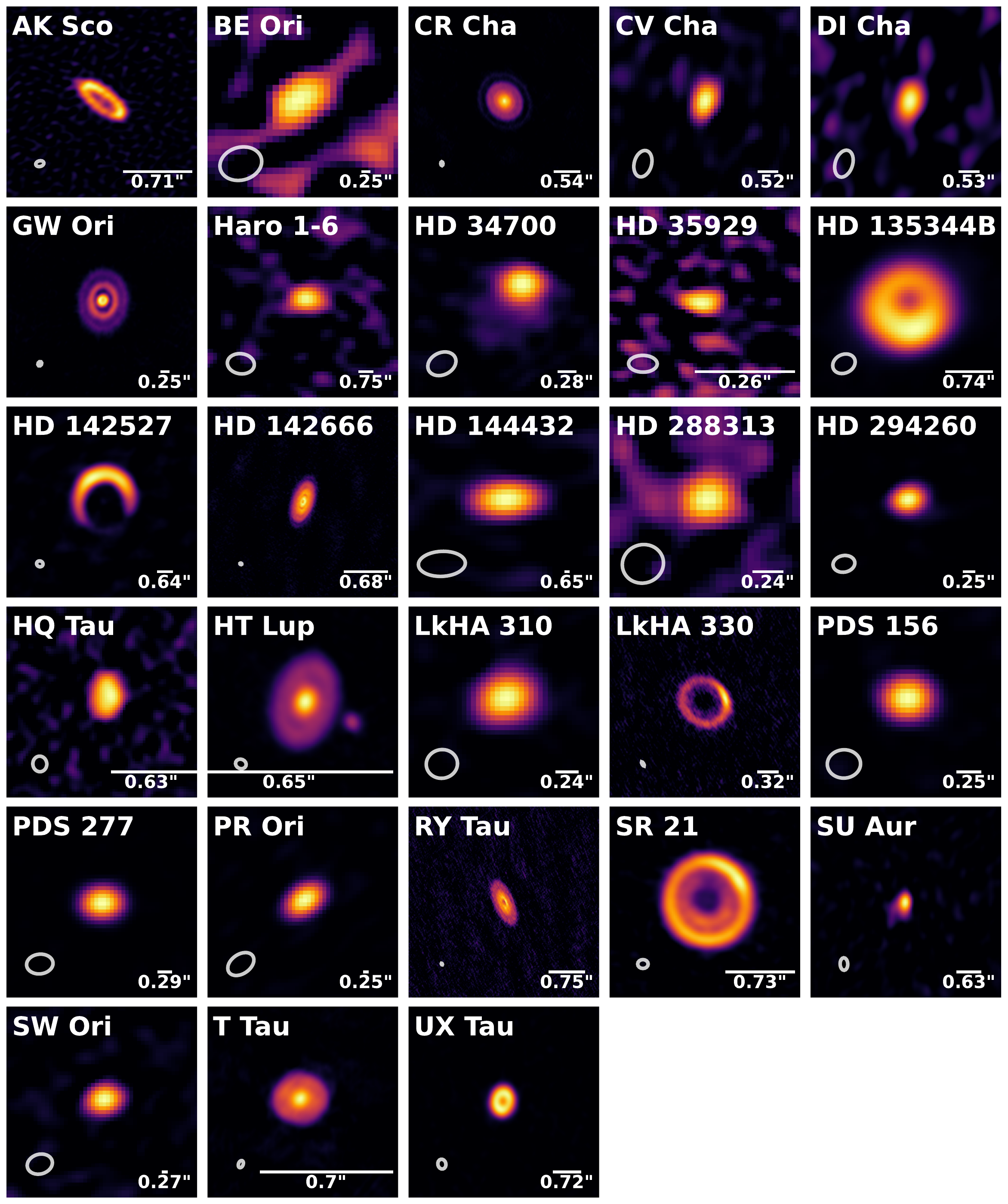}
    \caption{Continuum images of Intermediate Mass T~Tauri disks with a detection, using a sinh stretch. The size of the image is indicated by the bar of size 100~au on the bottom right of each panel with the corresponding angular size. The beam size is shown as the ellipse on the bottom left of each panel. The group classification can be found in Table~\ref{tab:disk_parameters}.}
    \label{fig:gallery_continuum}
\end{figure*}

\begin{figure*}[h!]
    \centering
    \includegraphics[width=0.8\textwidth]{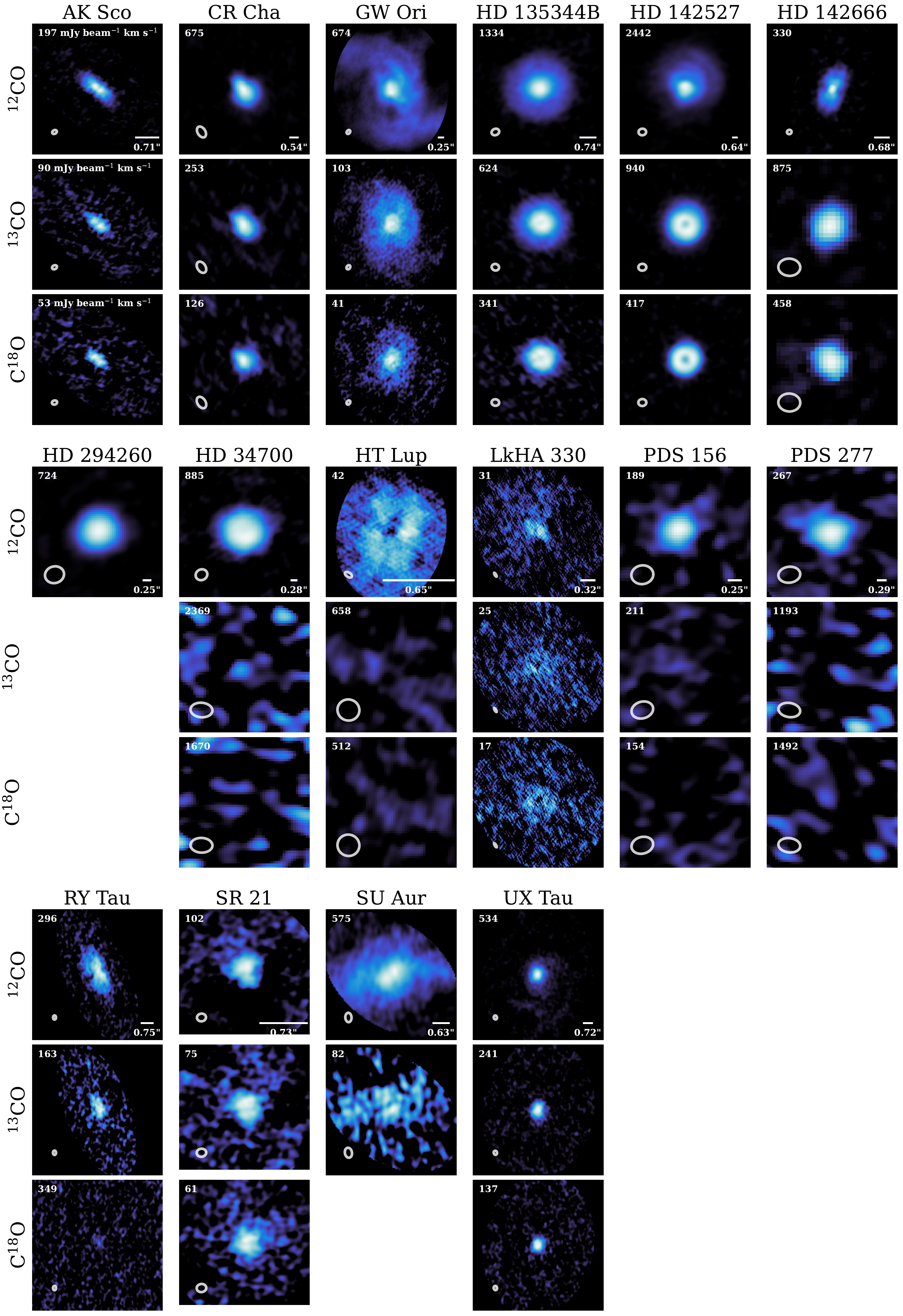}
    \caption{\ce{^12CO}, \ce{^13CO}, and \ce{C^18O} velocity integrated-intensity maps as obtained with Keplerian masking for all IMTTs which have a detection in at least one of the three isotopologues. An empty panel indicates that no observations are available of that particular isotopologue.}
    \label{fig:gallery_lines}
\end{figure*}

\section{Model setup}
\label{sec:model_setup}
For the modeling of the CO isotopologues, we use the \texttt{DALI} grid for Herbig disks presented in \citetalias{Stapper2024}. \texttt{DALI} \citep{Bruderer2012, Bruderer2013} is a thermo-chemical code which takes heating, cooling, and chemical processes into account and uses these to solve for the gas and dust thermal structure of the disk. The models used by \citetalias{Stapper2024} use the CO isotopologue chemistry network of \citet{Miotello2016}, which includes isotope-selective photodissociation, fractionation reactions, self-shielding, and freeze-out. The models use a parametric description for the density structure, motivated by a viscous accreting disk \citep{LyndenBell1974, Hartmann1998}. The vertical gas distribution is given by a Gaussian distribution. The grid run by \citetalias{Stapper2024} consists of models with disk masses ranging from $10^{-5}$ to $10^{-0.5}$~$M_\odot$, have different radial and vertical mass distributions, and different stellar luminosities and disk inclinations (see \citetalias{Stapper2024} for more details). The models were run until chemical equilibrium was reached at 1~Myr.

T~Tauri stars have lower effective temperatures compared to Herbig stars resulting in less UV photospheric emission in the former compared to the latter. As accretion contributes significantly to the overall UV budget, UV emission from accretion is added to the stellar spectrum for T~Tauri stars \citep[e.g.][]{Miotello2014, Miotello2016}. This is in contrast to Herbig stars, for which the accretion UV is not significant compared to the UV photons already emitted by the star itself \citep{Miotello2016}. But in the case of the IMTTs, accretion UV does significantly add to the overall UV budget.

To test this, we compare the grid of models from \citetalias{Stapper2024} to two smaller grids of models with the only difference being that an effective stellar temperature of 5500~K and luminosity of 10~$L_\odot$ are used for the IMTT disks based on the values of \citet{Valegard2021}, excluding the largest disk models from \citetalias{Stapper2024}. One grid has accretion UV added and one grid does not. The accretion luminosity is set as 1.5~$L_\odot$, the median value from the works of \citet{Calvet2004}, from modeling at wavelengths $>2000$~Å, and \citet{Wichittanakom2020}, from H$\alpha$ emission. The accretion was added to the stellar spectrum with a blackbody at 10$^4$~K. The addition of the accretion UV results in $L_\text{fuv}$/$L_\text{bol}=1.1\times10^{-2}$, compared to $L_\text{fuv}$/$L_\text{bol}=1.2\times10^{-3}$ without, and $L_\text{fuv}$/$L_\text{bol}=7.4\times10^{-2}$ for the Herbig disks. The addition of the accretion gives an order of magnitude difference compared to without, similar to the difference between the T~Tauri and Herbig disk models of \citet{Miotello2016}. Figure \ref{fig:IMTT_model_hists} compares the resulting disk integrated \ce{C^18O} luminosities of both grids to the Herbig disk grid from \citetalias{Stapper2024}, raytraced at a distance of 100~pc. Each panel presents a different gas mass, from $10^{-4}$ to $10^{-2}$ solar masses. The lower effective temperature of the IMTT stars decreases the UV luminosity compared to Herbig stars, which increases the \ce{C^18O} luminosity of the disk for the lowest disk masses due to a decrease in photo-dissociation of CO. As the CO gas becomes optically thin, either due to a larger disk or a lower disk mass, self-shielding becomes less efficient. For the highest disk mass considered in Fig.~\ref{fig:IMTT_model_hists} the disk integrated \ce{C^18O} luminosity is very similar to that of the Herbig disks. When including the accretion UV the \ce{C^18O} luminosity decreases, compensating for the decrease in UV emission due to the lower effective temperature. This effectively moves the distribution back again to that of the Herbig disks. This shows that the Herbig disk models can be used for the IMTT models and hence, we use the same models in this work as was used in \citetalias{Stapper2024}.

\begin{figure*}[t]
    \centering
    \includegraphics[width=\textwidth]{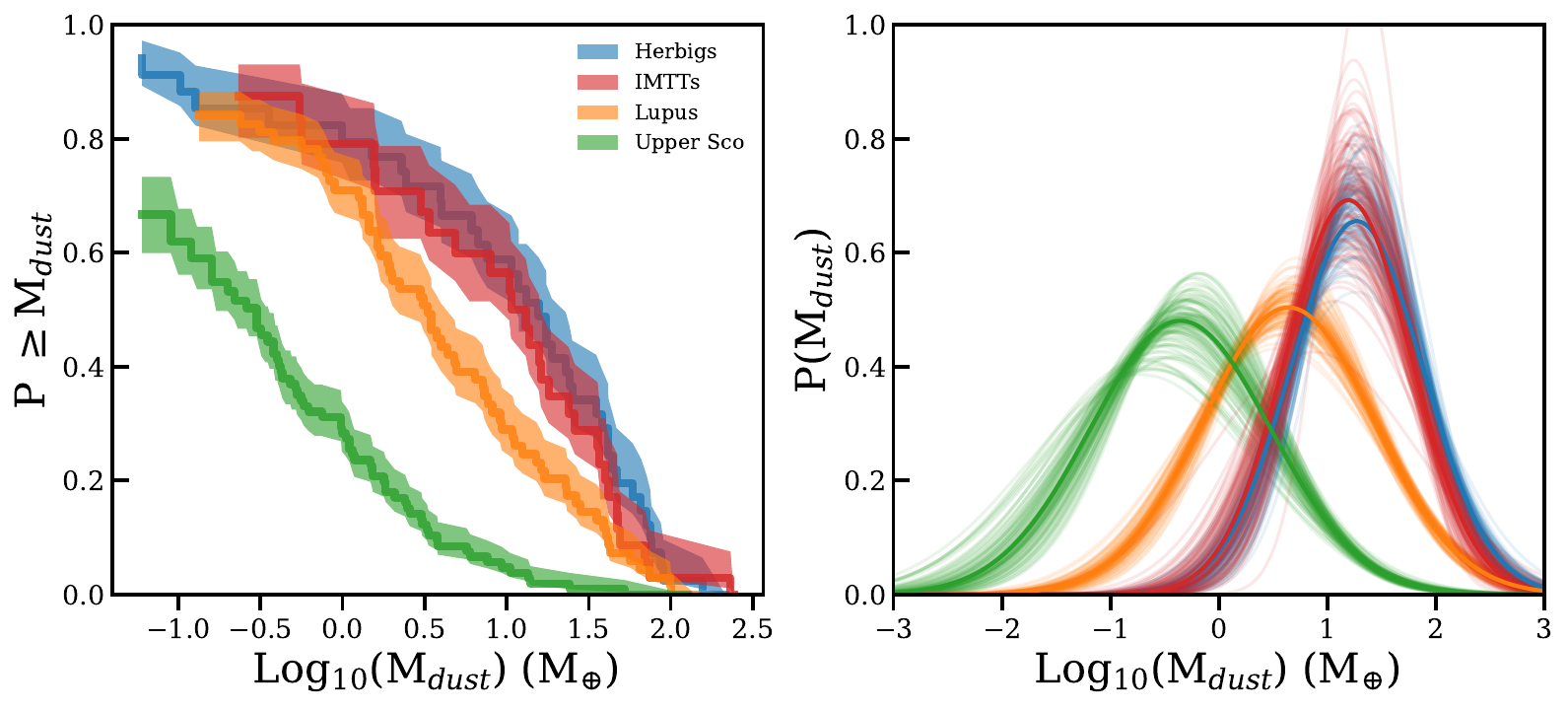}
    \caption{Dust mass distributions of the IMTTs, Herbig disks \citep{Stapper2022}, disks in Lupus \citep{Ansdell2016}, and disks in Upper~Sco \citep{Barenfeld2016}. The left panel shows the cumulative distributions, with the shaded region indicating the $1\sigma$ confidence interval. The right panel presents the probability distributions obtained by fitting a log-normal distribution to the cumulative distributions using a bootstrapping method. This figure shows a clear similarity between the IMTT and Herbig dust mass distributions.}
    \label{fig:cdf_dust}
\end{figure*}

\begin{figure}
    \centering
    \includegraphics[width=0.5\textwidth]{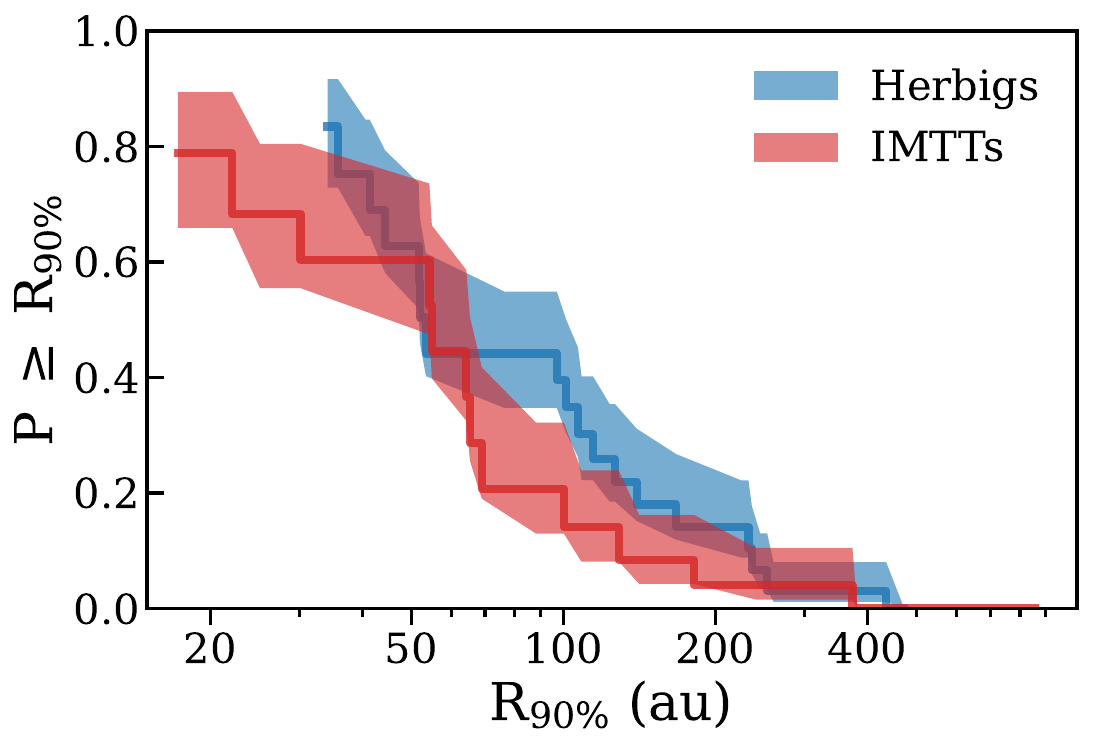}
    \caption{Cumulative distributions of the 90\% dust radii of the Herbig disks \citepalias{Stapper2024} and of the IMTTs. Both samples have a similar radius distribution.}
    \label{fig:cdf_radii}
\end{figure}

\section{Results}
\label{sec:results}
\subsection{Continuum images}
\label{subsec:continuum_images}
Figure \ref{fig:gallery_continuum} presents the continuum images of all IMTTs observed with ALMA. Similar to the observations of Herbig disks in \citetalias{Stapper2022}, there is a large variety of sizes and structures visible in our set of IMTT observations. The majority of disks are unresolved, but all disks which are resolved exhibit single or multiple rings. In total, 12 disks have resolved observations showing structures, out of the 26 detected disks, this is 46\%. This is lower compared to the fraction found by \citetalias{Stapper2024}, which was 60\% (15/25), which is likely due to the lower spatial resolution of the data available in our work, and this percentage will increase with higher resolution data. AK~Sco, HD~135344B, HD~142527, and HD~142666, which are also classified as Herbig disks, are part of these resolved disks. Removing these disks lowers the number of resolved IMTT disks to 8, showing a clear lack of deep observations towards IMTT disks.

Apart from disks with resolved structures, there are three unresolved disks which are noteworthy. For HD~35929 the data give strong constraints on the size of the dust disk, less than 41~au, while still being unresolved. Also SU~Aur and HD~34700, which show extended asymmetric emission in the dust, while the main disk is unresolved. Particularly SU~Aur is famous for its large arm in CO due to late-stage infall, which could be related to this asymmetry \citep{Ginski2021}. 

There are also multiple stellar systems present in our sample. HD~288313 in particular is a complex system of at least three components \citep{Reipurth2010}. The disk around the A component (the Herbig star) is quite faint in the continuum emission, while one of the other components has a peak flux an order of magnitude higher. HT~Lup is a triple system, with all three components having clear millimeter continuum emission \citep{Kurtovic2018}. The A component around the Herbig star has the largest disk. In Fig.~\ref{fig:gallery_continuum}, the B component is also visible, while the C component is outside of the field of view. Lastly, T~Tau is a triple stellar system \citep{Kohler2016}.

\begin{figure*}
    \centering
    \includegraphics[width=\textwidth]{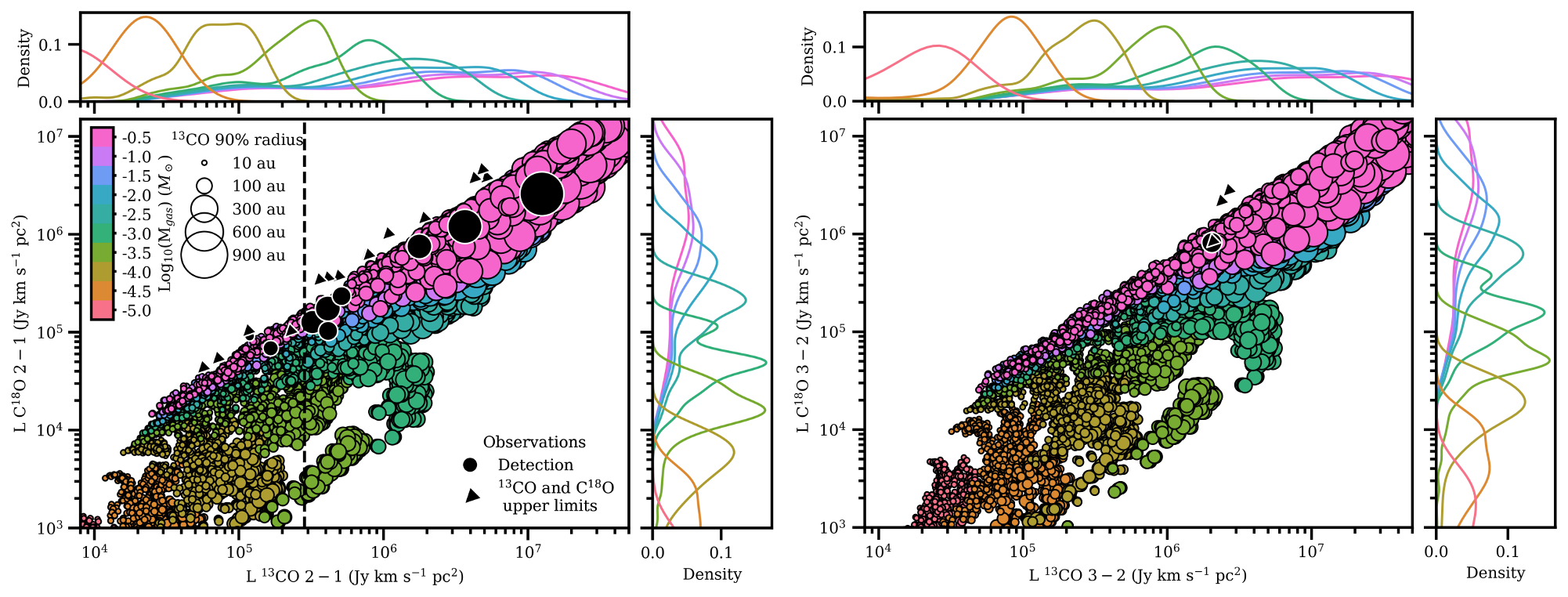}
    \caption{\ce{^13CO} and \ce{C^18O} disk integrated luminosities for transitions $J=2-1$ (left panel) and $J=3-2$ (right panel). The colors correspond to the DALI models from \citetalias{Stapper2024}, while the black markers correspond to the observations. The vertical line in the left panel indicates the \ce{^13CO} luminosity of SU~Aur, as no \ce{C^18O} observations are available.}
    \label{fig:13CO_vs_C18O}
\end{figure*}

\begin{figure*}
    \centering
    \includegraphics[width=\textwidth]{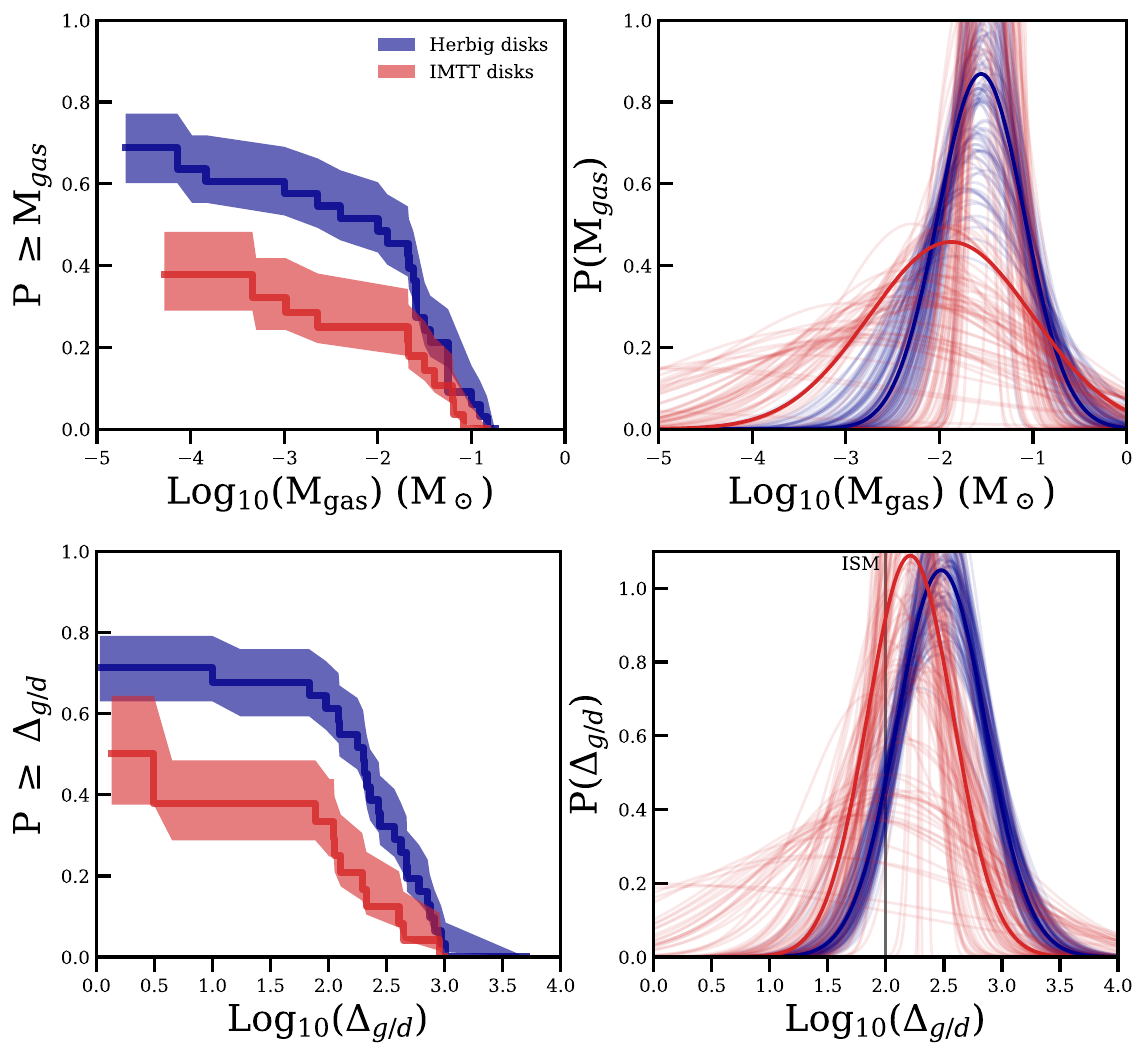}
    \caption{The resulting gas masses obtained from the models of \citetalias{Stapper2024} are shown in the top row. The Herbig disks are shown in blue, while the distribution of the IMTT disks is shown in red. Combining the gas distributions with the dust mass distributions in Fig.~\ref{fig:cdf_dust} results in the gas-to-dust ratio distributions in the bottom row.}
    \label{fig:cdf_gas}
\end{figure*}

\begin{figure}
    \centering
    \includegraphics[width=0.5\textwidth]{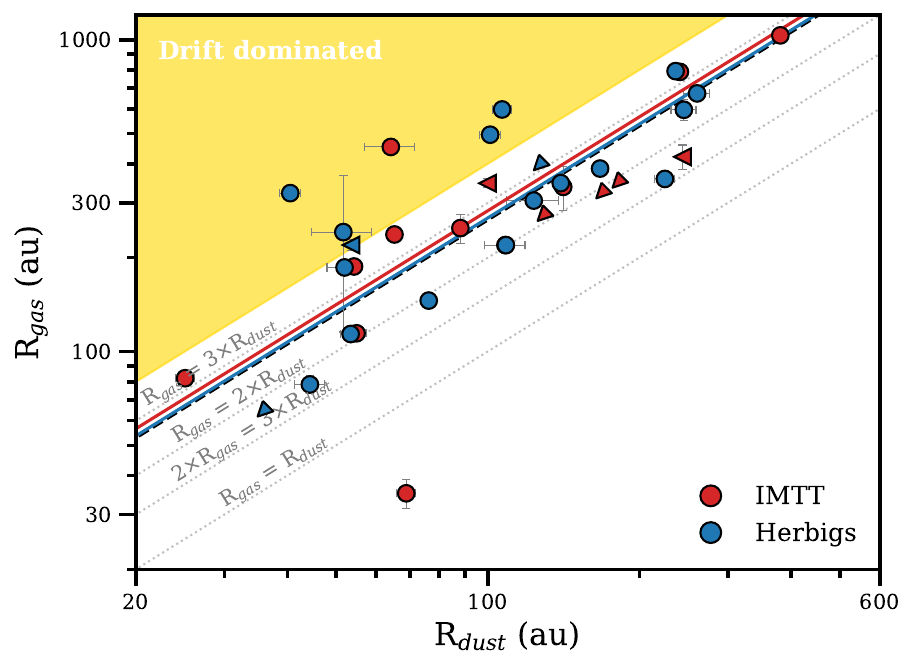}
    \caption{Dust and gas 90\% radii of the IMTT disks (red) and Herbig disks (blue). The fitted relation between the two parameters is shown as the solid lines. The relationship is the same for both Herbig and IMTT disks. In yellow the region is shown where the difference between the dust and gas radii cannot be explained by optical depth effects only \citep{Trapman2019}. The dashed black line is a fit through the IMTT disks excluding SR~21, GW~Ori, HT~Lup, and SU~Aur, as these have extended emission beyond the disk.}
    \label{fig:rgas_vs_rdust}
\end{figure}

\subsection{Gas images}
\label{subsec:gas_images}
The moment zero maps of disks for which at least one of the three CO isotopologues are detected are presented in Figure~\ref{fig:gallery_lines}. If one of the isotopologues is detected, the other non-detections are also shown by integrating over the same velocity range as is done for the detected isotopologue. If a panel is empty, no observations of that particular isotopologue are available. 

Again, a large variety in structures can be seen. In particular cavities are visible in three of our disks in both \ce{^13CO} and \ce{C^18O}: HD~135344B, LkH$\alpha$~330, and HD~142527 \citep[see for more on these disks e.g.,][]{vanderMarel2015, Brown2008, Pinilla2022, Temmink2023}. This is indicative of deep cavities, where these molecules become optically thin. Apart from cavities other structures are visible as well, in particular envelopes and streamers. These potential streamers and envelopes can be seen in the \ce{^12CO} emission of GW~Ori, T~Tau \citep{Rota2022}, and SU~Aur \citep{Ginski2021}. This extended emission does pose difficulties in obtaining a measure of the radii and integrated fluxes. HT~Lup has large unresolved structures visible by large fringe patterns in the observations, and strong continuum absorption in the center of the disk (see the cavity-like appearance in Fig.~\ref{fig:gallery_lines}). Other disks with absorption include SR~21 in particular, with strong absorption features on the south-western side of the disk resulting in an asymmetric appearance in \ce{^12CO}. 

Out of the 31 disks with at least one of the three CO isotopologues available, 16 out of 28 (57\%) for which \ce{^12CO} data are available have a detection, while this is 11 out of 29 (38\%) for \ce{^13CO}, and 10 out of 28 (36\%) for \ce{C^18O}. In general, if \ce{^13CO} is detected, then \ce{C^18O} is detected as well. On the other hand, there are four disks which have \ce{^12CO} detected, while the other CO isotopologues are not.

\subsection{Dust masses and radii}
\label{subsec:dust_masses}
We can obtain cumulative distributions from the dust masses using the \texttt{lifelines} package \citep{DavidsonPilon2021}, together with the probability distributions by fitting a log-normal distribution, see the left and right panel of Fig.~\ref{fig:cdf_dust} respectively. In addition, the cumulative distribution of the Herbig disks from \citetalias{Stapper2022} (with a median age of 5.5~Myr and a median stellar mass of 2~$M_\odot$), and the distributions of the disks in Lupus \citep[1-3~Myr, median stellar mass of 0.24~$M_\odot$,][]{Ansdell2016} and Upper~Sco \citep[5-10~Myr, median stellar mass of 0.26~$M_\odot$,][]{Barenfeld2016} are also shown.

The distributions of the Herbig dust masses and the IMTT dust masses stand out as very similar. At around 60~$M_\oplus$ the IMTT distribution does dip below the Herbig distribution, but is still within the shaded region indicating the $1\sigma$ confidence interval. Using a Kolmogorov-Smirnov test using \texttt{scipy} \citep{2020SciPy-NMeth}, we can assess whether the distributions are sampled from the same underlying distribution. We obtain a $p$-value of 0.962, indicating that we can accept the null hypothesis and that both distributions have the same underlying dust mass distribution.

Following \citetalias{Stapper2022}, we can obtain the probability distributions by fitting a log-normal distribution to the cumulative distributions using a bootstrapping method with $10^5$ samples. These fits result in the probability distributions shown in the right panel of Figure~\ref{fig:cdf_dust}. The resulting distributions show a clear overlap between the Herbig disks and IMTTs. Table \ref{tab:fit_params} shows the resulting means and standard deviations of the distributions. We include the dust masses of five Herbig disks\footnote{BH~Cep, BO~Cep, HD~200775, SV~Cep, XY~Per} observed with NOEMA from \citetalias{Stapper2024}, hence the difference in obtained parameters for the Herbig disk dust mass distribution compared to the one in \citetalias{Stapper2022}. Both the mean and the standard deviation of the Herbig disks and IMTTs fall within the given uncertainty intervals. Hence, further supporting the similarities between the two distributions.

Figure~\ref{fig:cdf_radii} presents the cumulative distributions of the $R_{90\%}$ dust radii for the IMTT disks and for the Herbig disks \citepalias{Stapper2024}. The radii are very similar between the two populations, as was also the case for the dust masses. The largest IMTT disk in our sample is the disk of GW~Ori, with a 90\% radius of 382~au. The other resolved disks range from 236~au down to 25~au with a median of 69~au. Again using a Kolmogorov-Smirnov test, we can test the null hypothesis of both distributions being sampled from the same populations. Indeed, we cannot reject this null hypothesis based on the resulting $p$-value of 0.860. Hence,the two distributions can be sampled from the same underlying population.

\renewcommand{\arraystretch}{1.2}
\begin{table}[t!]
\caption{Log-normal distribution fit results for the dust mass cumulative distributions shown in Fig. \ref{fig:cdf_dust}. The $M_\text{dust}$ parameters are given in log$_{10}$($M/M_\oplus$).}
\centering
\begin{tabular}{l|cc}
\hline\hline
          & $\mu$                   & $\sigma$               \\ \hline
Herbigs   & 1.27$^{+0.05}_{-0.05}$  & 0.61$^{+0.06}_{-0.06}$ \\
IMTTs     & 1.19$^{+0.06}_{-0.07}$  & 0.58$^{+0.09}_{-0.08}$ \\
Lupus     & 0.64$^{+0.04}_{-0.05}$  & 0.79$^{+0.05}_{-0.04}$ \\
Upper Sco & -0.36$^{+0.11}_{-0.14}$ & 0.83$^{+0.09}_{-0.07}$ \\ \hline
\end{tabular}\\
\label{tab:fit_params}
\end{table}



\subsection{Gas masses and radii}
\label{subsec:gas_masses}
In this section we compare the disk integrated luminosities of the \ce{^13CO} and \ce{C^18O} isotopologues to the DALI models obtained by \citetalias{Stapper2024}. This is shown in Figure~\ref{fig:13CO_vs_C18O}. In the left panel, the $J=2-1$ transition is shown, while the right panel shows the $J=3-2$ transition of the CO isotopologues, for both the observations and the models of \citetalias{Stapper2024}.

Comparing the observed and modeled \ce{^13CO} and \ce{C^18O} luminosities as shown in Fig.~\ref{fig:13CO_vs_C18O}, it is clear that for the detected disks, most are in the optically thick regime similar to the Herbig disks, as the smallest disks are on the bottom left, while the largest disks are on the top right. For the models, the multiple orders of magnitude in disk mass overlap with the same \ce{^13CO} and \ce{C^18O} fluxes, while for the larger disks the models fan out. The `hook'-shape for models with the same disk mass are due to an increased disk radius which reduces the self-shielding capacity of the CO molecules, reducing the emission of the \ce{C^18O} first and then the \ce{^13CO}. See \citetalias{Stapper2024} for more details.

For the disks in which the CO isotopologues are detected, we obtain the disk mass after selecting models based on the disk and stellar parameters in Table~\ref{tab:disk_parameters} (see \citetalias{Stapper2024} for details on this selection process). The cumulative gas mass and gas-to-dust ratio distributions and their corresponding probability distributions can be found in Fig.~\ref{fig:cdf_gas}. We find a gas mass distribution of Log$_{10}$($M_\text{disk}$)=-1.88±0.87~$M_\odot$ for the IMTTs, compared to Log$_{10}$($M_\text{disk}$)=-1.55±0.46~$M_\odot$ for the Herbigs. Importantly, while the upper limits for the Herbig disks with non-detections of \ce{^13CO} and \ce{C^18O} were constraining enough to obtain a limit on the disk mass, the upper limits obtained for the IMTT disks are not. Hence, most have upper limits of the maximum disk mass in the model grid. The obtained gas masses are in general 0.3~dex lower compared to the Herbig disks, see Figure~\ref{fig:cdf_gas}. The distribution is wider and goes towards lower gas masses compared to the Herbig disks.

Combining the gas and dust mass distributions of the IMTT and Herbig disks from Figs.~\ref{fig:cdf_dust} and \ref{fig:cdf_gas}, we obtain two similar gas-to-dust ratio distributions in the bottom row of Fig.~\ref{fig:cdf_gas}. Apart from RY~Tau, which has a relatively low \ce{C^18O} flux compared to its \ce{^13CO} flux, all disks show gas-to-dust ratios of at least the ISM value of 100 or higher, with a distribution of Log$_{10}$($\Delta_{g/d}$)=2.21±0.37 for the IMTTs compared to Log$_{10}$($\Delta_{g/d}$)=2.48±0.38 for the Herbigs. This is in line with the findings for the Herbig disks in \citetalias{Stapper2024}. The distribution of the IMTT disks seems to be lower by $\sim0.2$~dex, but this falls within the 1-sigma width of the distributions. Hence, we conclude that there is no indication of the IMTT gas-to-dust ratios being different when compared to the Herbig disks. The fact that high gas-to-dust ratios are found for these disks could be an indication of that the dust masses as determined via eq.~(\ref{eq:mdust}) are underestimated. This could indicate that either the assumption of optically thin emission does not hold, or that the assumed dust opacity is incorrect. Recent works have indeed shown that dust masses can be underestimated by up to a factor of $\sim7$ due to these effects \citep[e.g.,][]{Liu2022, Xin2023, Savvidou2024}.

The dust and gas \ce{^12CO} 90\% radii are presented in Fig.~\ref{fig:rgas_vs_rdust}, in which the radii are compared to those of the Herbig disks (\citetalias{Stapper2022} and \citetalias{Stapper2024}). The dust radial drift dominated limit from \citet{Trapman2019} is also shown, which indicates the region where optical depth effects cannot solely account for the difference seen between the gas and dust radius, hence radial drift is needed. Fitting a relationship through the scatter of the resolved disks we find a ratio of 2.8±0.2 between the gas and dust radius for the IMTT disks. This is the same compared to the ratio of the Herbig disks of 2.7±0.2 within the uncertainties. We note that for four IMTT disks the \ce{^12CO} radius is rather difficult to determine. GW~Ori has large spiral-like structures on the north and south side of the disk. As this is not the case for the \ce{^13CO} disk, the \ce{^12CO} disk is likely measured to be larger. HT~Lup has foreground contamination, but the disk is visible in \ce{^12CO}. Therefore a maximum size was set by eye, within which the 90\% radius was measured. SR~21 has absorption on the south side of the disk, reducing the size of the \ce{^12CO} disk. Still, for the other two isotopologues the disk is not found to be large, the CO isotopologues are only peaking inside the cavity of this transition disk. This results in a smaller gas disk size compared to the dust disk size. Lastly, due to the infalling streamer, the size of the SU~Aur disk in \ce{^12CO} is likely larger than it would be without the streamer. Removing these four disks results in a gas to dust radius ratio of 2.6±0.3, again the same as that for the Herbig disks. This also coincides with a factor of 2.5 found by \citet{Andrews2020}.

\section{Discussion}
\label{sec:discussion}

\begin{figure}[t]
    \centering
    \includegraphics[width=0.5\textwidth]{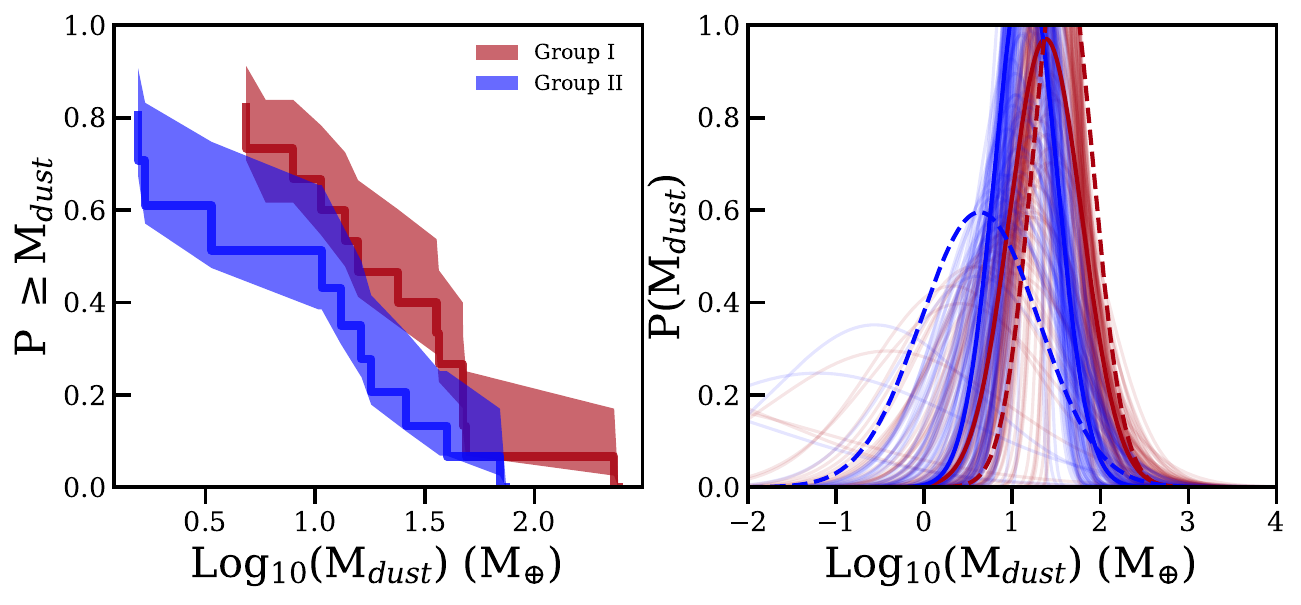}
    \caption{Dust cumulative distributions of the group~I and group~II disks. There is no significant difference, in contrast to what is found for the Herbig disks \citepalias{Stapper2022}. The dotted distributions in the right panel are the dust mass distributions for the Herbig disks from \citetalias{Stapper2022}.}
    \label{fig:cdf_dust_groups}
\end{figure}

\subsection{Evolution of Herbig disks}
\label{subsec:evolution}
As IMTTs are the precursors of Herbig stars, the fact that the disk dust radii and masses are the same may be unexpected, as an evolution towards smaller less massive disks due to dust evolutionary processes is expected. As proposed in \citetalias{Stapper2022}, the stopping of radial drift due to massive exoplanets can explain the large difference seen in the mass and size of Herbig disks when compared to T~Tauri disks. As no difference in the radius or the dust mass is found between the Herbig disks and the IMTT disks, this may indicate that massive exoplanets have already formed in the IMTT disks stopping radial drift, resulting in a higher inferred disk mass. As both the radii and dust mass distributions are indistinguishable for both IMTTs and Herbigs, this implies that planets are forming early in their lifetime, dominating the (dust) evolution of these disks, giving very similar results. Furthermore, by selecting stars with infrared excess with masses between 1.5 and 3.5 solar masses within 300~pc, \citet{Iglesias2022} found that most of these pre-main sequence intermediate mass stars are already evolved towards the debris disk stage in less than 10~Myr (though the lack of accretion in these stars may indicate that these disks are older). Given the stark similarities between Herbig and IMTT disks, this separation into full disks and debris disks should happen early on in the disk lifetime, and might highly depend on planet formation happening in these disks. Therefore, an important step forward would be to obtain a sample of even younger intermediate mass stars, possibly even earlier classes.

In \citetalias{Stapper2022}, a dust mass dichotomy was found between the group~I and group~II disks as defined by \citet{Meeus2001}. These are defined as the SED of group~I disks can be reproduced by a power-law and a black-body, while the SED of group~II disks can be reconstructed with a power-law alone. In particular the distribution of dust masses of the group~II disks was very similar to the dust mass distribution of the disks in Lupus \citep{Ansdell2016}. On the other hand, the group~I disks were more massive. Moreover, in contrast to the mostly full disks of the group~II disks, group~I disks were found to have large inner cavities. As \citet{Valegard2021} has classified the IMTTs into group~I and group~II based on the \citet{Meeus2001} classification, we can ascertain if the same differences between the two groups are found.

Figure~\ref{fig:cdf_dust_groups} presents the two cumulative mass distributions in the left panel, with the fitted log-normal probability distributions in the right panel. The results from this fit can be found in Table~\ref{tab:fit_params_dust_groups}. We do not find a clear difference between the two populations, which was found for the Herbig disks \citepalias{Stapper2022}. However, we do note that the maximum and minimum dust masses are indeed lower for the group~II disks compared to the group~I disks. Comparing the distributions in the left panel of Fig.~\ref{fig:cdf_dust_groups} with the distributions presented by \citetalias{Stapper2022}, the mean dust mass of the group~I disks is the same, the main difference is in the group~II disks.

As was noted by \citet{Honda2012}, \citet{Maaskant2013}, and \citetalias{Stapper2022}, group~I disks tend to be transitional disks, having large inner cavities, while group~II disks tend to have full disks. With the IMTTs this is the case as well. For the resolved disks, the group~I disks indeed show cavities (AK~Sco, SR~21, HD~135344B, HD~142527, LkH$\alpha$~330, and UX~Tau), and the group~II disks indeed have full disks (CR~Cha, GW~Ori, HD~142666, HT~Lup, and RY~Tau). We note that the classification of AK~Sco by \citet{Valegard2021} is group~I, while it is classified as group~II by \citet{GuzmanDiaz2021} as used by \citetalias{Stapper2022}. Hence, the ALMA observations are of particular necessity to characterize the disk. The difference in mass and morphology between the group~I and group~II disks was interpreted by \citetalias{Stapper2022} as the group~I disks forming massive exoplanets which create an inner cavity by stopping radial drift, while the group~II disks most of the solid mass drifts inwards reducing the total inferred disk mass. As there is no difference in the inferred dust mass for the group~I disks when comparing the Herbig disks with the IMTT disks, no further evolution has happened in these particular disks, which is expected if the radial drift has stopped. On the other hand, the group~II Herbig disks have lower disk masses than the group~II IMTT disks, which is expected if the radial drift has not been stopped in the disk. So, most disks around intermediate-mass pre-main sequence stars converge quickly (i.e., within the timescale of the ages of the IMTT and Herbig disks) to small/compact disks unless prevented by a massive exoplanet. This should occur well before the age of the IMTT disks, i.e., 5~Myr. We are mainly tracing the survivors of this process, as a large fraction of the pre-main sequence intermediate mass stars are like those of \citet{Iglesias2022}, without a disk, or the intermediate mass equivalent of the Naked T~Tauri stars. A key question to consider is whether this apparent quick convergence towards small, or absent, disks around pre-main sequence intermediate mass stars is a natural phenomenon or related to biases ingrained in how these stellar populations are defined and/or observed. This on/off behavior is not observed in T~Tauri disks, for which there is a clear decrease in dust mass over time \citep{Drazkowska2023}. This might be related to the need for high accretion rates to be classified as a Herbig star, as discussed in \citet{GrantStapper2023}. Moreover, the addition of mostly observing the brightest disks may bias this even further. A complete volume-limited millimeter survey of pre-main sequence intermediate mass stars is therefore highly needed (Stapper et al. in prep.).

This hypothesis is also supported by observations of metallicities of the Herbig stars themselves \citep{Kama2015, GuzmanDiaz2023}. Stars with a group~I disk generally have lower metallicities compared to the group~II disks, which is associated with the presence of massive exoplanets accreting the refractory elements and stopping the radial drift of the dust inwards. \citet{Brittain2023} point out that there is evidence that the opacities and temperatures of the dust are not the same between the two groups \citep{Woitke2019}. Group~I disks tend to have smaller grains compared to group~II disks, resulting in a higher inferred disk mass in the former compared to the latter. That no differences between the group~I and group~II disks are found in Fig.~\ref{fig:cdf_dust_groups} might be an effect of similar dust populations in the younger IMTT disks and that this differentiates when evolving towards a Herbig disk. This must be further investigated using longer wavelength observations.

Lastly, recent work by \citet{Vioque2023} has shown that intermediate-mass young stellar objects become significantly less clustered with time. Hence, the IMTT disks are expected to be more clustered than Herbig stars. Given the fact that we do not find significant differences between both populations might indicate that the environment does not play a major role in the evolution of disks around intermediate mass stars. A study dedicated to the effect of the environment is therefore highly needed.

\renewcommand{\arraystretch}{1.2}
\begin{table}[t!]
\caption{Log-normal distribution fit results for the dust mass cumulative distributions shown in Fig. \ref{fig:cdf_dust_groups}. The $M_\text{dust}$ parameters are given in log$_{10}$($M/M_\oplus$).}
\centering
\begin{tabular}{l|cc}
\hline\hline
          & $\mu$                   & $\sigma$               \\ \hline
Group I   & 1.38$^{+0.13}_{-0.25}$  & 0.41$^{+0.23}_{-0.15}$ \\
Group II  & 1.15$^{+0.08}_{-0.15}$  & 0.35$^{+0.19}_{-0.14}$ \\ \hline
\end{tabular}\\
\label{tab:fit_params_dust_groups}
\end{table}

\begin{figure}
    \centering
    \includegraphics[width=0.5\textwidth]{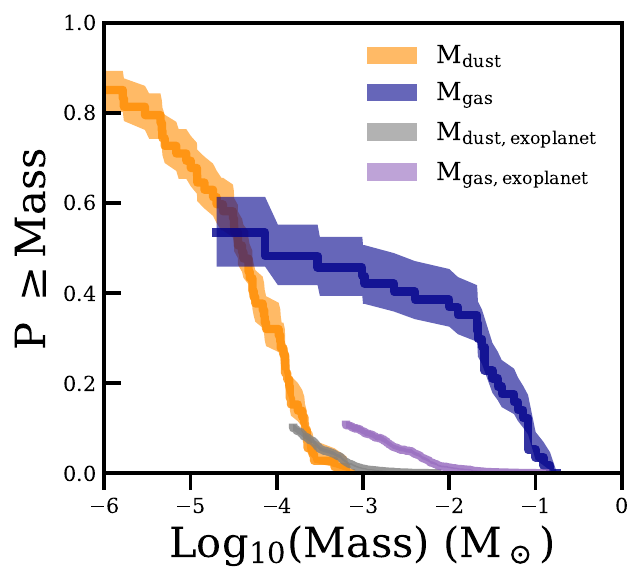}
    \caption{Comparing the dust and gas mass cumulative distributions of the disks around intermediate mass (IMTT disks + Herbig disks) with a dust and gas mass distribution of exoplanets.}
    \label{fig:cdf_exoplanet}
\end{figure}

\subsection{Planet formation around intermediate mass stars}
\label{subsec:planet_formation}
The dust masses of planet-forming disks have been directly compared to the solid mass of massive exoplanets in planetary systems by \citet{Tychoniec2020}. They found a clear discrepancy between class~II disks and the masses present in massive exoplanets, which would not be able to form. Even when taking into account observational biases this discrepancy would only not be a problem when the formation efficiency would be 100\% \citep{Mulders2021}, which is likely not the case \citep[see for an overview][]{Drazkowska2023}. As this comparison only has been done based on the dust mass inferred from continuum observations, this section will also directly compare the exoplanetary masses to the gas masses we found in the disks around intermediate mass  stars. Moreover, we also include observational biases into the distribution.

To obtain an updated exoplanet mass distribution, we use the NASA exoplanet archive\footnote{\url{https://exoplanetarchive.ipac.caltech.edu}}, and select all confirmed exoplanets. Following \citet{Tychoniec2020}, we set no detection method and use both the planet mass $M_p$ and minimum planet mass $M_p\sin(i)$ in our distribution. To obtain a distribution for the intermediate mass stars, we only include the planets around stars with a minimum mass of 1.5~$M_\odot$. To obtain a distribution taking the observational biases into account, we use the work of \citet{Wolthoff2022}, who determined giant exoplanet occurrence rates around stars with masses higher than 0.8~$M\odot$. Based on this work, we normalize the distribution to 1.3\% for planetary masses higher than 9~$M_\text{jup}$, to 5.7\% for masses higher than 2.7~$M_\text{jup}$ and to 15.8\% for the complete distribution. We note that planets with a period ranging from 80 to 3600 days are used here, so there are differences in the scales traced by planets and disks. Following \citet{Tychoniec2020}, the total planetary mass includes both the core mass and the atmosphere, so we use the relationship from \citet{Thorngren2016} to obtain a heavy-element mass of the planet, and we assume that this is the dust mass necessary to form the planet. The total gas (atmosphere) mass is then simply the difference between the planetary mass and the heavy-element mass of the planet. After the core and atmosphere masses, all planetary systems belonging to the same star are combined into a single mass. The number of stars hosting at least one giant planet is 10.7\% \citep{Wolthoff2022}, hence we again normalize the distribution so that it adds up to 10.7\%. The resulting distribution is shown in Fig.~\ref{fig:cdf_exoplanet}.

Figure~\ref{fig:cdf_exoplanet} also shows the dust mass and gas mass distributions after combining the data of our work with the dust masses from \citetalias{Stapper2022} and the gas masses from \citetalias{Stapper2024}, after removing the duplicate disks from the sample of our work. The tail end of dust mass distribution of the disks is similar to the dust masses necessary to build the exoplanetary systems. However, an efficiency of 100\% is necessary. This is in line with what \citet{Mulders2021} finds. Comparing the exoplanet dust mass distribution with the disk gas mass distribution, assuming a gas-to-dust ratio of 100, the lower mass planets may be less of a problem to still form. Still, it is clear that the cores of the massive exoplanets should already have formed in these systems. This is in line with the previous discussion, as these exoplanets are thought to have a big impact on the evolution of disks around intermediate mass stars.

Comparing the gas mass distributions, there still is more than enough gas mass available in the disks to form the exoplanet atmosphere. For a large fraction of the population there is at least one or two orders of magnitude difference in the mass available in the disks compared to what is necessary to form the exoplanets. Hence, planet envelope accretion might still be ongoing, while the cores of the planets have already been formed.

\section{Conclusion}
\label{sec:conclusion}
In this work we analyze the continuum, \ce{^12CO}, \ce{^13CO}, and \ce{C^18O} emission of 35 Intermediate Mass T~Tauri (IMTT) disks, all being ALMA archival data. The obtained dust masses and radii are compared to those of Herbig disks \citep{Stapper2022}, and T~Tauri disks \citep{Barenfeld2016, Ansdell2016}. From our results the following can be concluded:

\begin{enumerate}
    \item IMTT disks have both the same dust mass distribution, assuming optically thin emission, and dust radius distribution as Herbig disks. Hence, a similar difference in dust mass is found when compared to the T~Tauri disks.
    \item The gas mass of IMTT disks, as determined with DALI thermochemical models, is possibly slightly lower than that of Herbig disks, but this is likely due to the lack in sensitivity of the available CO isotopologue observations in the archive to obtain meaningful gas mass limits. Deeper observations of IMTT disks are urgently needed.
    \item The gas radii are the same as that of the Herbig disks. Compared to the dust radii, the same ratio between the two is found as for the Herbig disks.
    \item Dividing the IMTT disks into group~I (rising FIR slope) and group~II (decreasing FIR slope) disks reveals no significant difference regarding the dust mass. This is in stark contrast to what is found for Herbig disks, for which the inferred dust mass in the group~II disks is lower than for the group~I disks. This difference between the Herbig group~I and group~II disks might be indicative of different evolutionary scenarios happening in these two groups. Group~I disks stay the same due to a massive exoplanet stopping radial drift, while the group~II disks rapidly shrink over time due to radial drift decreasing the inferred disk mass.
    \item As the mass and sizes of IMTT disks are the same as for Herbig disks, planet formation has likely already started in these disks, shaping their formation, and subsequent evolution towards Herbig disks. Most disks around intermediate-mass pre-main sequence stars converge quickly to small disks unless prevented by a massive exoplanet. We are mainly tracing the survivors of this process, as most pre-main sequence intermediate mass stars are debris disks.
    \item Comparing the disk dust mass distributions to the amount of dust mass in massive exoplanets, assuming that this equals the heavy-element mass of the planet, it is clear that the cores of the exoplanets already need to have formed, as there is not enough dust mass present in the disks. However, comparing the disk gas mass distributions to the mass in planetary envelopes, there is more than an order of magnitude difference. This indicates that while the core of the exoplanets may already have formed, they are likely still accreting their envelope.
    \end{enumerate}

IMTTs are an important class of objects for understanding planet formation and would benefit from targeted, high resolution deep imaging observations with ALMA in the future.

\begin{acknowledgements}
The research of LMS is supported by the Netherlands Research School for Astronomy (NOVA). The project leading to this publication has received support from ORP, that is funded by the European Union's Horizon 2020 research and innovation programme under grant agreement No 101004719 [ORP]. This paper makes use of the following ALMA data: 2012.1.00158.S, 2012.1.00313.S, 2012.1.00870.S, 2013.1.00426.S, 2013.1.00437.S, 2013.1.00498.S, 2013.1.01075.S, 2015.1.00222.S, 2015.1.01353.S, 2015.1.01600.S, 2016.1.00204.S, 2016.1.00484.L, 2016.1.00545.S, 2016.1.01164.S, 2016.1.01338.S, 2017.1.00286.S, 2017.1.01353.S, 2017.1.01460.S, 2018.1.00689.S, 2018.1.01302.S, 2019.1.00703.S, 2019.1.00951.S, 2019.1.01813.S, 2021.2.00005.S, 2021.1.00854.S, 2021.1.01705.S, 2022.1.01155.S, 2022.1.01460.S . ALMA is a partnership of ESO (representing its member states), NSF (USA) and NINS (Japan), together with NRC (Canada), MOST and ASIAA (Taiwan), and KASI (Republic of Korea), in cooperation with the Republic of Chile. The Joint ALMA Observatory is operated by ESO, AUI/NRAO and NAOJ. This research has made use of the NASA Exoplanet Archive, which is operated by the California Institute of Technology, under contract with the National Aeronautics and Space Administration under the Exoplanet Exploration Program. We would like to thank Allegro for their support and computing resources for reducing and imaging the ALMA data. Also, we would like to thank Gijs Mulders for his useful comments on the exoplanet mass distribution. We also thank the referee for their careful consideration of our work and for their thoughtful comments which improved the manuscript. This work makes use of the following software: The Common Astronomy Software Applications (CASA) package \citep{McMullin2007}, Dust And LInes \citep[DALI,][]{Bruderer2012, Bruderer2013}, Python version 3.9, astropy \citep{astropy2013, astropy2018}, lifelines \citep{DavidsonPilon2021}, matplotlib \citep{Hunter2007}, numpy \citep{Harris2020}, scipy \citep{2020SciPy-NMeth} and seaborn \citep{Waskom2021}.

\end{acknowledgements}

\bibliographystyle{aa}
\bibliography{references.bib}

\appendix


\section{Datasets used}
\label{app:data_sets}
In Table \ref{tab:project_codes} the project codes of the used data sets are listed together with their spatial and velocity resolution and the (line-free) rms noise.

\onecolumn
\renewcommand{\arraystretch}{1.2}
\begin{landscape}
{\fontsize{7}{8.4}\selectfont  
\begin{longtable}[c]{l|cc|ccc|ccc|ccc|r}
\caption{Data sets and corresponding parameters for each IMTT disk. The rms noise is for an empty channel at the given velocity resolution with units mJy~beam$^{-1}$. 
\label{tab:project_codes}
}\\
\hline\hline
& \multicolumn{2}{c|}{Continuum} & \multicolumn{3}{c|}{\ce{^12CO}} & \multicolumn{3}{c|}{\ce{^13CO}} & \multicolumn{3}{c|}{\ce{C^18O}} & \\ \hline
\makecell{Name \\ \hspace{1mm}} & \makecell{Spat.res. \\ ($''$)} & \makecell{rms \\ \hspace{1mm}} & \makecell{Spat.res. \\ ($''$)} & \makecell{Vel.res. \\ (km~s$^{{-1}}$)} & \makecell{rms \\ \hspace{1mm}} & \makecell{Spat.res. \\ ($''$)} & \makecell{Vel.res. \\ (km~s$^{{-1}}$)} & \makecell{rms \\ \hspace{1mm}} & \makecell{Spat.res. \\ ($''$)} & \makecell{Vel.res. \\ (km~s$^{{-1}}$)} & \makecell{rms \\ \hspace{1mm}} & \makecell{Project codes} \\
\hline
\endfirsthead
\caption{Continued.}\\
\hline\hline
& \multicolumn{2}{c|}{Continuum} & \multicolumn{3}{c|}{\ce{^12CO}} & \multicolumn{3}{c|}{\ce{^13CO}} & \multicolumn{3}{c|}{\ce{C^18O}} & \\ \hline
\makecell{Name \\ \hspace{1mm}} & \makecell{Spat.res. \\ ($''$)} & \makecell{rms \\ \hspace{1mm}} & \makecell{Spat.res. \\ ($''$)} & \makecell{Vel.res. \\ (km~s$^{{-1}}$)} & \makecell{rms \\ \hspace{1mm}} & \makecell{Spat.res. \\ ($''$)} & \makecell{Vel.res. \\ (km~s$^{{-1}}$)} & \makecell{rms \\ \hspace{1mm}} & \makecell{Spat.res. \\ ($''$)} & \makecell{Vel.res. \\ (km~s$^{{-1}}$)} & \makecell{rms \\ \hspace{1mm}} & \makecell{Project codes} \\
\hline
\endhead
\hline
\endfoot
AK Sco & 0.09 $\times$ 0.06 (-76\degree) & 0.05 & 0.15 $\times$ 0.11 (-63\degree) & 1.00 & 3.55 & 0.16 $\times$ 0.12 (-63\degree) & 1.00 & 3.97 & 0.16 $\times$ 0.13 (-75\degree) & 1.00 & 2.50 & 2016.1.00204.S \\
Ass ChaT2-21 & 0.48 $\times$ 0.31 (1\degree) & 0.39 & 0.55 $\times$ 0.38 (-3\degree) & 0.42 & 23.27 &  &  &  &  &  &  & 2012.1.00313.S$^{*}$\hspace{-4.4pt} \\
Ass ChaT2-54 & 0.76 $\times$ 0.40 (45\degree) & 0.87 & 0.81 $\times$ 0.43 (45\degree) & 0.40 & 75.75 & 0.74 $\times$ 0.56 (-42\degree) & 0.40 & 435.53 & 0.76 $\times$ 0.47 (-14\degree) & 0.40 & 429.18 & 2013.1.01075.S$^{*}$\hspace{-4.4pt} \\
BE Ori & 1.33 $\times$ 1.03 (-75\degree) & 0.31 & 1.60 $\times$ 1.19 (-73\degree) & 0.20 & 44.40 & 1.75 $\times$ 1.25 (-69\degree) & 0.20 & 46.02 & 1.75 $\times$ 1.26 (-71\degree) & 0.20 & 39.70 & 2019.1.00951.S \\
Brun 656 & 0.94 $\times$ 0.66 (-69\degree) & 0.83 &  & & &  & & &  & & & 2017.1.01353.S \\
CO Ori$^{*}$ & 7.08 $\times$ 4.83 (84\degree) & 1.78 & 7.62 $\times$ 5.21 (78\degree) & 0.20 & 282.44 & 7.86 $\times$ 5.43 (81\degree) & 0.20 & 284.19 & 7.87 $\times$ 5.47 (83\degree) & 0.20 & 225.54 & 2021.2.00005.S \\
CR Cha & 0.08 $\times$ 0.05 (5\degree) & 0.02 & 0.72 $\times$ 0.47 (32\degree) & 0.20 & 10.57 & 0.75 $\times$ 0.49 (33\degree) & 0.20 & 12.06 & 0.76 $\times$ 0.49 (32\degree) & 0.20 & 8.70 & 2017.1.00286.S \\
CV Cha & 0.71 $\times$ 0.45 (-15\degree) & 3.70 &  & & & 0.73 $\times$ 0.56 (-44\degree) & 0.20 & 542.70 & 0.75 $\times$ 0.48 (-14\degree) & 0.20 & 544.64 & 2013.1.00437.S$^{*}$\hspace{-4.4pt} \\
DI Cha & 0.73 $\times$ 0.45 (-18\degree) & 3.35 &  & & & 0.70 $\times$ 0.48 (-7\degree) & 0.30 & 300.56 & 0.69 $\times$ 0.48 (-8\degree) & 0.30 & 513.42 & 2013.1.00437.S$^{*}$\hspace{-4.4pt} \\
GW Ori & 0.15 $\times$ 0.12 (-29\degree) & 0.08 & 0.20 $\times$ 0.15 (-37\degree) & 0.20 & 5.85 & 0.21 $\times$ 0.16 (-36\degree) & 0.20 & 6.75 & 0.21 $\times$ 0.15 (-37\degree) & 0.20 & 4.77 & 2017.1.00286.S \\
Haro 1-6 & 1.42 $\times$ 1.06 (83\degree) & 0.54 & 1.58 $\times$ 1.19 (86\degree) & 0.20 & 101.78 & 1.65 $\times$ 1.24 (86\degree) & 0.20 & 110.71 & 1.65 $\times$ 1.23 (84\degree) & 0.20 & 62.65 & 2016.1.00545.S \\
HBC 442$^{*}$ & 7.79 $\times$ 4.38 (83\degree) & 2.11 & 8.23 $\times$ 4.90 (81\degree) & 0.20 & 308.71 & 8.54 $\times$ 5.04 (80\degree) & 0.20 & 312.10 & 8.57 $\times$ 5.17 (80\degree) & 0.20 & 271.11 & 2021.2.00005.S \\
HBC 502 & 1.77 $\times$ 1.06 (-80\degree) & 5.50 &  & & &  & & &  & & & 2016.1.01338.S \\
HD 34700$^{*}$ & 0.46 $\times$ 0.35 (-64\degree) & 0.19 & 0.54 $\times$ 0.48 (-63\degree) & 0.63 & 10.05 &  &  &  &  &  &  & 2021.1.01705.S \\
 &  &  &  &  &  & 7.75 $\times$ 5.12 (85\degree) & 0.20 & 296.24 & 7.77 $\times$ 5.20 (88\degree) & 0.20 & 227.99 & 2021.2.00005.S \\
HD 35929 & 0.08 $\times$ 0.04 (89\degree) & 0.03 & 0.11 $\times$ 0.06 (-62\degree) & 0.20 & 4.86 & 0.11 $\times$ 0.07 (-69\degree) & 0.35 & 3.21 & 0.11 $\times$ 0.07 (-71\degree) & 0.35 & 2.62 & 2021.1.00854.S \\
HD 135344B & 0.37 $\times$ 0.29 (-63\degree) & 0.98 & 0.36 $\times$ 0.29 (-68\degree) & 0.20 & 29.59 &  &  &  &  &  &  & 2012.1.00870.S$^{*}$\hspace{-4.4pt} \\
           &  &  &  &  &  & 0.34 $\times$ 0.30 (75\degree) & 0.20 & 19.06 & 0.35 $\times$ 0.30 (84\degree) & 0.20 & 25.93 & 2012.1.00158.S$^{*}$\hspace{-4.4pt} \\
HD 142527 & 0.27 $\times$ 0.24 (69\degree) & 0.76 & 0.93 $\times$ 0.81 (-85\degree) & 0.20 & 16.11 & 0.97 $\times$ 0.84 (-87\degree) & 0.20 & 16.76 & 0.98 $\times$ 0.85 (-88\degree) & 0.20 & 11.76 & 2015.1.01353.S \\
HD 142666 & 0.03 $\times$ 0.02 (65\degree) & 0.03 & 0.21 $\times$ 0.20 (-70\degree) & 0.32 & 5.83 &  &  &  &  &  &  & 2016.1.00484.L \\
  &  &  &  &  &  & 1.05 $\times$ 0.82 (87\degree) & 0.20 & 42.43 & 1.04 $\times$ 0.85 (87\degree) & 0.20 & 27.06 & 2015.1.01600.S \\
HD 144432 & 7.39 $\times$ 3.94 (-87\degree) & 1.89 & 7.53 $\times$ 4.12 (-88\degree) & 1.50 & 92.27 & 7.98 $\times$ 4.37 (-90\degree) & 1.50 & 67.76 & 8.01 $\times$ 4.38 (-90\degree) & 1.50 & 54.38 & 2022.1.01460.S \\
HD 288313 & 0.33 $\times$ 0.30 (-72\degree) & 0.25 & 0.39 $\times$ 0.36 (-65\degree) & 0.64 & 10.16 & 0.42 $\times$ 0.39 (-69\degree) & 0.66 & 11.89 & 0.42 $\times$ 0.39 (-68\degree) & 0.67 & 10.09 & 2022.1.01155.S \\
HD 294260 & 0.46 $\times$ 0.35 (-79\degree) & 0.22 & 0.58 $\times$ 0.51 (-75\degree) & 0.63 & 7.87 &  & & &  & & & 2021.1.01705.S \\
HQ Tau & 0.12 $\times$ 0.11 (6\degree) & 0.12 &  & & & 0.15 $\times$ 0.13 (0\degree) & 0.20 & 20.81 & 0.15 $\times$ 0.14 (4\degree) & 0.20 & 15.78 & 2016.1.01164.S \\
HT Lup & 0.03 $\times$ 0.03 (68\degree) & 0.02 & 0.07 $\times$ 0.04 (57\degree) & 0.32 & 1.25 &  &  &  &  &  &  & 2016.1.00484.L \\
       &  &  &  &  &  & 0.29 $\times$ 0.28 (74\degree) & 0.40 & 33.54 & 0.29 $\times$ 0.28 (70\degree) & 0.40 & 24.82 & 2015.1.00222.S \\
LkHA 310 & 0.33 $\times$ 0.30 (-81\degree) & 0.17 & 0.40 $\times$ 0.36 (-70\degree) & 0.64 & 12.13 & 0.43 $\times$ 0.39 (-71\degree) & 0.66 & 10.22 & 0.42 $\times$ 0.38 (-70\degree) & 0.67 & 9.48 & 2022.1.01155.S \\
LkHA 330 & 0.08 $\times$ 0.03 (30\degree) & 0.02 & 0.10 $\times$ 0.03 (32\degree) & 1.50 & 1.26 & 0.10 $\times$ 0.03 (31\degree) & 1.50 & 1.33 & 0.11 $\times$ 0.03 (31\degree) & 1.50 & 1.18 & 2018.1.01302.S \\
PDS 156 & 0.35 $\times$ 0.30 (-87\degree) & 0.15 & 0.41 $\times$ 0.35 (-82\degree) & 0.63 & 11.79 & 0.49 $\times$ 0.37 (-74\degree) & 0.66 & 11.59 & 0.49 $\times$ 0.37 (-75\degree) & 0.67 & 10.47 & 2022.1.01155.S \\
PDS 277$^{*}$ & 0.56 $\times$ 0.41 (-84\degree) & 0.24 & 0.70 $\times$ 0.51 (-83\degree) & 0.63 & 12.28 &  &  &  &  &  &  & 2021.1.01705.S \\
        &  &  &  &  &  & 7.13 $\times$ 4.58 (81\degree) & 0.20 & 209.84 & 7.09 $\times$ 4.59 (83\degree) & 0.20 & 188.04 & 2021.2.00005.S \\
PR Ori & 1.52 $\times$ 1.01 (-55\degree) & 0.70 &  & & &  & & &  & & & 2019.1.01813.S \\
RY Ori$^{*}$ & 7.76 $\times$ 4.47 (86\degree) & 2.16 & 8.35 $\times$ 5.03 (83\degree) & 0.20 & 308.84 & 8.66 $\times$ 5.15 (83\degree) & 0.20 & 312.46 & 8.68 $\times$ 5.28 (82\degree) & 0.20 & 263.91 & 2021.2.00005.S \\
RY Tau & 0.04 $\times$ 0.02 (27\degree) & 0.05 &  &  &  &  &  &  &  &  &  & 2017.1.01460.S \\
       &  &  & 0.25 $\times$ 0.18 (-1\degree) & 0.70 & 15.36 & 0.26 $\times$ 0.19 (0\degree) & 0.70 & 13.61 & 0.27 $\times$ 0.19 (0\degree) & 0.70 & 9.91 & 2013.1.00498.S \\
SR 21 & 0.11 $\times$ 0.09 (90\degree) & 0.09 & 0.14 $\times$ 0.12 (-83\degree) & 0.35 & 7.01 & 0.15 $\times$ 0.12 (-83\degree) & 0.35 & 7.37 & 0.15 $\times$ 0.12 (-82\degree) & 0.35 & 5.68 & 2018.1.00689.S \\
SU Aur & 0.33 $\times$ 0.20 (2\degree) & 0.19 & 0.36 $\times$ 0.23 (3\degree) & 0.40 & 14.81 & 0.38 $\times$ 0.27 (10\degree) & 0.40 & 16.69 &  & & & 2013.1.00426.S \\
SW Ori & 1.34 $\times$ 1.03 (-74\degree) & 0.31 & 1.61 $\times$ 1.18 (-73\degree) & 0.20 & 44.77 & 1.75 $\times$ 1.25 (-69\degree) & 0.20 & 45.80 & 1.76 $\times$ 1.26 (-70\degree) & 0.20 & 40.45 & 2019.1.00951.S \\
T Tau & 0.04 $\times$ 0.03 (-21\degree) & 0.08 &  &  &  &  &  &  &  &  &  & 2019.1.00703.S \\
UX Tau & 0.26 $\times$ 0.21 (12\degree) & 0.08 & 0.30 $\times$ 0.25 (10\degree) & 0.32 & 8.44 & 0.31 $\times$ 0.26 (13\degree) & 0.66 & 6.19 & 0.32 $\times$ 0.26 (12\degree) & 0.67 & 4.71 & 2013.1.00498.S \\
\end{longtable}
\tablefoot{Disks with an asterisk have been observed with the ACA, all others with ALMA.}}

\end{landscape}
\renewcommand{\arraystretch}{1.0}

\end{document}